\documentclass[pdflatex,sn-standardnature]{sn-jnl} % 
\jyear{2023}%

\theoremstyle{thmstyleone}%

\theoremstyle{thmstyletwo}%

\theoremstyle{thmstylethree}%

\usepackage[version=4]{mhchem} % chem package
\usepackage{lineno}
\usepackage{comment} % commenting out whole sections
\usepackage{astjnlabbrev}
\usepackage{subfigure}
\usepackage{graphicx}
\usepackage{makecell}
\usepackage{rotating}
\usepackage{multibib} %[resetlabels]
\newcites{App}{Methods References}

\usepackage{xcolor}
\begin{document}
\title[Mid-Infrared Spectrum of WASP-39b]{Sulphur dioxide in the mid-infrared transmission spectrum of WASP-39b}
% Paper leads
\author*[1,2]{Diana Powell}\email{diana.powell@uchicago.edu\textmd{\\\textcolor{black}{All author affiliations are listed at the end of the paper.}}}
\author[2,3]{Adina D. Feinstein}
\author[4]{Elspeth K. H. Lee}
% Team leads (reverse alphabetical + Peter)
\author[2,5]{Michael Zhang}
\author[6]{Shang-Min Tsai}
\author[7,8]{Jake Taylor}
\author[9]{James Kirk}
\author[10,11]{Taylor Bell}
\author[12]{Joanna K. Barstow}
\author[13]{Peter Gao}
% Tier 1 
\author[2]{Jacob L. Bean}
\author[14,15]{Jasmina Blecic}
\author[16]{Katy L. Chubb}
\author[17]{Ian J. M. Crossfield}
\author[18]{Sean Jordan}
\author[3]{Daniel Kitzmann}
\author[19]{Sarah E. Moran}
\author[20,21,22]{Giuseppe Morello}
\author[23]{Julianne I. Moses}
\author[24,25]{Luis Welbanks}
\author[26]{Jeehyun Yang}
\author[27]{Xi Zhang}
% Tier 2

\author[28,29]{Eva-Maria Ahrer}
\author[30]{Aaron Bello-Arufe}
\author[17]{Jonathan Brande}
\author[31]{S. L. Casewell}
\author[32]{Nicolas Crouzet}
\author[33,34]{Patricio E. Cubillos}
\author[4,35]{Brice-Olivier Demory}
\author[36]{Achr\`ene Dyrek}
\author[37]{Laura Flagg}
\author[30,38]{Renyu Hu}
\author[38]{Julie Inglis}
\author[4]{Kathryn D. Jones}
\author[39]{Laura Kreidberg}
\author[1]{Mercedes L\'opez-Morales}
\author[36]{Pierre-Olivier Lagage}
\author[4]{Erik A. Meier Vald\'es}
\author[32,40]{Yamila Miguel}
\author[41]{Vivien Parmentier}
\author[13]{Anjali A. A. Piette}
\author[42,43,5]{Benjamin V.\ Rackham}
\author[8]{Michael Radica}
\author[44]{Seth Redfield}
\author[45]{Kevin B. Stevenson}
\author[46]{Hannah R. Wakeford}

% Tier 3
\author[47]{Keshav Aggarwal}
\author[13]{Munazza K. Alam}
\author[48]{Natalie M. Batalha}
\author[49]{Natasha E. Batalha}
\author[8]{Bj\"{o}rn Benneke}
\author[50]{Zach K. Berta-Thompson}
\author[51]{Ryan P. Brady}
\author[52,53,54]{Claudio Caceres}
\author[48]{Aarynn L. Carter}
\author[55]{Jean-Michel D\'esert}
\author[56]{Joseph Harrington}
\author[57]{Nicolas Iro}
\author[24]{Michael R. Line}
\author[58]{Joshua D. Lothringer}
\author[59,25]{Ryan J. MacDonald}
\author[60,33,39]{Luigi Mancini}
\author[61,62]{Karan Molaverdikhani}
\author[48]{Sagnick Mukherjee}
\author[63]{Matthew C. Nixon}
\author[30]{Apurva V. Oza}
\author[21]{Enric Palle}
\author[64]{Zafar Rustamkulov}
\author[64,65]{David K. Sing}
\author[39]{Maria E. Steinrueck}
\author[66]{Olivia Venot}
\author[28,29]{Peter J. Wheatley}
\author[51]{Sergei N. Yurchenko}

\affil[1]{Center for Astrophysics ${\rm \mid}$ Harvard {\rm \&} Smithsonian, Cambridge, USA}
\affil[2]{Department of Astronomy \& Astrophysics, University of Chicago, Chicago, IL, USA}
\affil[3]{Laboratory for Atmospheric and Space Physics, University of Colorado Boulder, UCB 600, Boulder, CO 80309}
\affil[4]{Center for Space and Habitability, University of Bern, Bern, Switzerland}
\affil[5]{51 Pegasi b fellow}
\affil[6]{Department of Earth Sciences, University of California, Riverside, California, USA}
\affil[7]{Department of Physics, University of Oxford, Oxford, UK}
\affil[8]{Institut Trottier de Recherche sur les Exoplan\`etes and D\'epartement de Physique, Universit\'e de Montr\'eal, Montr\'eal, QC, Canada}
\affil[9]{Department of Physics, Imperial College London, London, UK}
\affil[10]{Bay Area Environmental Research Institute, NASA Ames Research Center, Moffett Field, CA, USA}
\affil[11]{Space Science and Astrobiology Division, NASA Ames Research Center, Moffett Field, CA, USA}
\affil[12]{School of Physical Sciences, The Open University, Milton Keynes, UK}
\affil[13]{Earth and Planets Laboratory, Carnegie Institution for Science, Washington, DC, USA}
\affil[14]{Department of Physics, New York University Abu Dhabi, Abu Dhabi, UAE}
\affil[15]{Center for Astro, Particle and Planetary Physics (CAP3), New York University Abu Dhabi, Abu Dhabi, UAE}
\affil[16]{Centre for Exoplanet Science, University of St Andrews, St Andrews, UK}
\affil[17]{Department of Physics \& Astronomy, University of Kansas, Lawrence, KS, USA}
\affil[18]{Institute of Astronomy, University of Cambridge, Cambridge, UK}
\affil[19]{Lunar and Planetary Laboratory, University of Arizona, Tucson, AZ, USA}
\affil[20]{Department of Space, Earth and Environment, Chalmers University of Technology, Gothenburg, Sweden}
\affil[21]{Instituto de Astrof\'isica de Canarias (IAC), Tenerife, Spain}
\affil[22]{INAF--Palermo Astronomical Observatory, Piazza del Parlamento, Palermo, Italy}
\affil[23]{Space Science Institute, Boulder, CO, USA}
\affil[24]{School of Earth and Space Exploration, Arizona State University, Tempe, AZ, USA}
\affil[25]{NHFP Sagan Fellow}
\affil[26]{Planetary Sciences Section, Jet Propulsion Laboratory, California Institute of Technology, Pasadena, USA}
\affil[27]{Department of Earth and Planetary Sciences, University of California Santa Cruz, Santa Cruz, California, USA}

\affil[28]{Centre for Exoplanets and Habitability, University of Warwick, Coventry, UK}
\affil[29]{Department of Physics, University of Warwick, Coventry, UK}
\affil[30]{Astrophysics Section, Jet Propulsion Laboratory, California Institute of Technology, Pasadena, CA, USA}
\affil[31]{School of Physics and Astronomy, University of Leicester, Leicester}
\affil[32]{Leiden Observatory, University of Leiden, Leiden, The Netherlands}
\affil[33]{INAF - Turin Astrophysical Observatory, Pino Torinese, Italy}
\affil[34]{Space Research Institute, Austrian Academy of Sciences, Graz, Austria}
\affil[35]{Space and Planetary Sciences, Institute of Physics, University of Bern}
\affil[36]{Universit\'e Paris-Saclay, CEA, CNRS, AIM, Gif-sur-Yvette, France}
\affil[37]{Department of Astronomy and Carl Sagan Institute, Cornell University, Ithaca, NY, USA}
\affil[38]{Division of Geological and Planetary Sciences, California Institute of Technology, Pasadena, CA, USA}
\affil[39]{Max Planck Institute for Astronomy, Heidelberg, Germany}
\affil[40]{SRON Netherlands Institute for Space Research, Leiden, the Netherlands}
\affil[41]{Universit\'e C\^ote d'Azur, Observatoire de la C\^ote d'Azur, CNRS, Laboratoire Lagrange, France}
\affil[42]{Department of Earth, Atmospheric and Planetary Sciences, Massachusetts Institute of Technology, Cambridge, MA, USA}
\affil[43]{Kavli Institute for Astrophysics and Space Research, Massachusetts Institute of Technology, Cambridge, MA, USA}
\affil[44]{Astronomy Department and Van Vleck Observatory, Wesleyan University, Middletown, CT, USA}
\affil[45]{Johns Hopkins Applied Physics Laboratory, Laurel, MD, USA}
\affil[46]{School of Physics, University of Bristol, Bristol, UK}
\affil[47]{Indian Institute of Technology, Indore, India}
\affil[48]{Department of Astronomy \& Astrophysics, University of California, Santa Cruz, Santa Cruz, CA, USA}
\affil[49]{NASA Ames Research Center, Moffett Field, CA, USA}
\affil[50]{Department of Astrophysical and Planetary Sciences, University of Colorado, Boulder, CO, USA}
\affil[51]{Department of Physics and Astronomy, University College London, United Kingdom}
\affil[52]{Instituto de Astrofisica, Facultad Ciencias Exactas, Universidad Andres Bello, Santiago, Chile}
\affil[53]{Centro de Astrofisica y Tecnologias Afines (CATA), Casilla 36-D, Santiago, Chile}
\affil[54]{Nucleo Milenio de Formacion Planetaria (NPF), Chile}
\affil[55]{Anton Pannekoek Institute for Astronomy, University of Amsterdam, Amsterdam, The Netherlands}
\affil[56]{Planetary Sciences Group, Department of Physics and Florida Space Institute, University of Central Florida, Orlando, Florida, USA}
\affil[57]{Institute of Planetary research, German aerospace center (DLR), Berlin, Germany}
\affil[58]{Department of Physics, Utah Valley University, Orem, UT, USA}
\affil[59]{Department of Astronomy, University of Michigan, Ann Arbor, MI, USA}
\affil[60]{Department of Physics, University of Rome \''Tor Vergata'', Rome, Italy}
\affil[61]{Universit\"ats-Sternwarte, Ludwig-Maximilians-Universit\"at M\"unchen, M\"unchen, Germany}
\affil[62]{Exzellenzcluster Origins, Garching, Germany}
\affil[63]{Department of Astronomy, University of Maryland, College Park, MD, USA}
\affil[64]{Department of Earth and Planetary Sciences, Johns Hopkins University, Baltimore, MD, USA}
\affil[65]{Department of Physics \& Astronomy, Johns Hopkins University, Baltimore, MD, USA}
\affil[66]{Universit\'e de Paris Cit\'e and Univ Paris Est Creteil, CNRS, LISA, Paris, France}

\abstract{\textbf{The recent inference of sulphur dioxide (\ce{SO2}) in the atmosphere of the hot ($\sim$1100 K), Saturn-mass exoplanet WASP-39b from near-infrared JWST observations \citep{jtec2023,Alderson2023,rustamkulov2023} suggests that photochemistry is a key process in high temperature exoplanet atmospheres \citep{Tsai2023}. This is due to the low ($<$1\,ppb) abundance of \ce{SO2} under thermochemical equilibrium, compared to that produced from the photochemistry of \ce{H2O} and \ce{H2S} (1-10\,ppm) \citep{Zahnle09,Zahnle2016,Hobbs2021,Tsai2021,Polman2022,Tsai2023}. However, the \ce{SO2} inference was made from a single, small molecular feature in the transmission spectrum of WASP-39b at 4.05 $\mu$m, and therefore the detection of other \ce{SO2} absorption bands at different wavelengths is needed to better constrain the \ce{SO2} abundance. Here we report the detection of \ce{SO2} spectral features at 7.7 and 8.5 $\mu$m in the 5--12\,$\mu$m transmission spectrum of WASP-39b measured by the JWST Mid-Infrared Instrument (MIRI) Low Resolution Spectrometer (LRS) \citep{kendrew15}. Our observations suggest an abundance of \ce{SO2} of 0.5--25\,ppm (1$\sigma$ range), consistent with previous findings \citep{Tsai2023}. In addition to \ce{SO2}, we find broad water vapour absorption features, as well as an unexplained decrease in the transit depth at wavelengths longer than 10\,$\mu$m. Fitting the spectrum with a grid of atmospheric forward models, we derive an atmospheric heavy element content (metallicity) for WASP-39b of $\sim$7.1--8.0 $\times$ solar and demonstrate that photochemistry shapes the spectra of WASP-39b across a broad wavelength range.}}

\maketitle

We observed WASP-39b using JWST MIRI/LRS on UTC 2023-02-14 from 15:03:20 to 22:59:36, spanning a total of 7.94 hours (Director’s Discretionary Time PID 2783). The observation included the full 2.8-hour transit, as well as 3 hours before and 1.87 hours after the transit to measure the stellar baseline. We used the slitless prism mode with no dithering. In this mode, MIRI/LRS yields a spectral range from 5--12\,$\mu$m, at an average resolving power of $R \equiv \lambda / \Delta \lambda \approx 100$, where $\lambda$ is the wavelength. The time-series observations included 1779 integrations of 16 seconds (100 groups per integration). No region of the detector was saturated.

We extracted the time-series stellar spectra using three independently developed reduction pipelines to test the impact of background modelling, spectral extraction method and aperture width, and light-curve-fitting routines on the resulting planetary transmission spectrum (see Methods and Extended Data Figures 1 and 2). We summed across the extracted stellar spectra to create white-light curves (Extended Data Figure 2) as well as binned spectrophotometric light curves for each pipeline (Figure 1). The light curves show clear instrumental systematics at the beginning of the observation that are driven by a decreasing exponential ramp effect \citep{bouwman2023}. At the detector level, the observations showed correlations with spatial position and an odd--even effect from row to row due to the readout time \citep{ressler15}. We do not see evidence of a very sharp, strong change in the initial exponential ramp's sign, amplitude, or timescale, known as a ``shadowed region", in our observations \cite[Extended Data Figure 1;][]{bell2023_arxiv}. We use wide spectrophotometric light curve bins of $\Delta \lambda = 0.25\mu$m to average over the odd--even row effect \citep{bell2023_arxiv} and we note that our conclusions are insensitive to the chosen bin size (smaller bins of 0.15 $\mu$m derive the same results) as well as the choice of the origin binning wavelength. 

We present the resulting transmission spectrum from each pipeline in Figure 2. Within the spectra, we are able to identify two broad absorption features belonging to \ce{SO2} at 7.7 and 8.5\,$\mu$m, which correspond to the asymmetric $\nu_3$ and symmetric $\nu_1$ fundamental bands, respectively, consistent with predictions from photochemical models \citep{Tsai2023}. We are also able to discern \ce{H2O} absorption, although it is mostly apparent between 5 and 7 $\mu$m owing to the overlapping \ce{SO2} feature at longer wavelengths. There is an abrupt decrease in the transit depth at $\lambda = 10$\,$\mu$m. The shadowed region systematic occurs from $\lambda \geq 10.6$ --- 11.8$\mu$m \citep{bell2023_arxiv}, at longer wavelengths compared to the abrupt decrease in the transmission spectrum.  
Therefore, if this abrupt change arose from the instrument and is not of astrophysical origin, then it is most likely driven by a different source of detector noise or an artifact that is not currently well understood.

In order to determine the detection significance of \ce{SO2} in our data and constrain its abundance, we conducted seven independent Bayesian retrievals on each of the three data reductions. Each nominal retrieval includes \ce{SO2} and \ce{H2O} as spectrally active gases, as well as a variety of cloud and haze treatments to account for degeneracies between retrieved cloud/haze properties and molecular abundances (see Methods). Other spectrally active gases were initially tested by the retrievals, including \ce{CH4}, \ce{NH3}, \ce{HCN}, \ce{CO}, \ce{CO2}, \ce{C2H2}, \ce{H2S}, but none of them showed significant detections. As shown in Figure 3 and Extended Data Table 4, the fits of the retrieval models to the data are generally good, with reduced chi-squared values close to 1. \ce{SO2} is detected to at least $\sim$3$\sigma$ significance for all retrieval frameworks and data reductions, except for one single retrieval--data reduction combination with a 2.5$\sigma$ detection, where other free parameters slightly reduced the SO$_2$ detection significance (see Methods). We retrieve a range of log volume mixing ratios from -6.3 to -4.6 (0.5--25\,ppm; lowest to highest 1$\sigma$ uncertainty bounds across all 6 retrieval frameworks) for the \texttt{Eureka!} reduction. 
Retrievals for the other reductions yielded similar results and are discussed in Methods and shown in Extended Data Figure 4.

Similar to \ce{SO2}, the retrieved \ce{H2O} abundances are largely consistent across all retrievals and reductions (see Extended Data Table 4 and Extended Data Figure 4), although the spread of values for the detection significance is greater than for \ce{SO2}, with some reduction-retrieval combinations yielding $\lesssim$2$\sigma$ while for others it is above 5$\sigma$. This serves to highlight the impact of choices made at both the reduction and retrieval stages on conclusions drawn from a spectrum. We postulate that the variation in detection significance that we see is due to the fact that the \ce{H2O} feature present in this observation is fairly broad, and likely impacted by the stronger \ce{SO2} feature at longer wavelengths and modelled haze properties at shorter wavelengths. For the Aurora/\texttt{Eureka!} 
combination the water abundance is relatively poorly constrained, with long tails in the distribution towards lower abundances and haze compensating for the relative lack of \ce{H2O} absorption at short wavelengths. Across the other six retrievals for the \texttt{Eureka!} reduction, the retrieved range of log volume mixing ratios is from -2.4 to -1.2 (0.4--6.3\%; lowest to highest 1$\sigma$ uncertainty).  

In addition to \ce{SO2} and \ce{H2O}, one retrieval framework found weak-to-moderate (2.5$\sigma$) evidence for \ce{SO}, with a feature between 8 and 10 $\mu$m (see Methods), which is predicted to be present by photochemical models \citep{Zahnle09,Tsai2023}, but additional observations would be needed to confirm or rule out its existence. Furthermore, we can largely rule out a grey cloud extending to low pressures with broad terminator coverage (see Methods), but more detailed cloud and haze properties such as particle sizes and cloud top pressure cannot be consistently constrained.     

We use a suite of independent forward-model grids that include photochemistry to infer the atmospheric metallicity and elemental ratios of WASP-39b from the observed \ce{SO2} abundance (see Methods). As \ce{SO2} is photochemical in origin, a rigorous treatment of photochemistry is vital for connecting \ce{SO2} to bulk atmospheric properties. Figure 4 shows the comparison between four independent photochemical models, all of which include moderately different chemical networks for H, C, O, N, and S molecules and use the same average atmospheric temperature--pressure profiles (morning and evening terminators), eddy diffusion profile, and stellar spectrum of WASP-39 adopted by ref.\,\citep{Tsai2023} as inputs. The model transmission spectra generated from the four photochemical models are largely consistent with each other and the data, showing that sufficient \ce{SO2} is generated photochemically to explain the 7.7 and 8.5\,$\mu$m absorption features. In particular, the limb-averaged volume mixing ratio of \ce{SO2} for the best-fitting 7.5$\times$ solar metallicity models span the range of 2.5--6.1 ppm, in line with our free-retrieval results (Extended Data Table 4). The 8.5 $\mu$m \ce{SO2} feature is notably sensitive to metallicity in this range while the strongest 7.7 $\mu$m feature starts to saturate with metallicity $\gtrsim$ 7.5 $\times$ solar. 

Using an expanded grid of one of the photochemical models \cite[see Methods;][]{crossfield:2023} we find best-fitting atmospheric metallicity values of 7.1--8.0 $\times$ solar across the three data reductions, as well as a consistent -- though weak -- preference for a super-solar O/S ratio, sub-solar C/O, and approximately solar C/S. Even though no carbon species is detected in the spectrum, constraints on the carbon abundance are still possible through the high degree of coupling between the CHONS elements in the photochemistry. These results are largely corroborated by comparisons to independent, self-consistent, radiative-convective-thermochemical equilibrium model grids that are post-processed to include \ce{SO2} (see Methods), which also infer a sub-solar C/O, as well as slightly higher atmospheric metallicity values ranging between 10--30$\times$ solar, depending on the specific data reduction. These findings are within the range of C/O (subsolar) and atmospheric metallicities (supersolar) derived from near-infrared JWST transmission spectra of WASP-39b using self-consistent radiative-convective thermal equilibrium grid models  \citep{jtec2023,feinstein2023,ahrer2023,Alderson2023,rustamkulov2023} and photochemical models that were able to match the near-infrared \ce{SO2} feature \citep{Tsai2023}. Our work therefore shows that JWST's MIRI LRS is fully capable of producing information-rich exoplanet observations like the near-infrared instruments.

The interpretation of WASP-39b's transmission spectrum at wavelengths beyond 10 $\mu$m is uncertain. If the observed sudden drop in transit depth is astrophysical in origin rather than due to an artifact in the data, then several possibilities exist. For example, the transit radius of a planet can decrease quickly with increasing wavelength when a cloud layer becomes sufficiently optically thin such that we can probe below the cloud base \citep{Vahidinia2014}. In addition, spectral features associated with the vibrational modes of bonds of several cloud and haze species are situated in the mid-infrared \citep{Wakeford2015,Gao2021,miles2023}, but none of the known features can explain our data.  Meanwhile, the absorption cross sections of some gaseous species, such as metal hydrides (e.g. \ce{SiH} and \ce{BeH}), can exhibit downward slopes starting at $\sim$10\,$\mu$m \citep{Tennyson2018}. However, the abundances of these species needed to explain the observed feature ($\sim$1000 ppm) are orders of magnitude greater than what is expected in a near-solar metallicity atmosphere (see Methods). Additional observations will be needed to explore the behavior and provenance of the $>$10 $\mu$m transmission spectrum of WASP-39b.

\bibliographystyle{sn-standardnature}
\bibliography{main.bib} 

\hfill \break

\begin{figure}[hbt!]
\centering
\includegraphics[width=0.99\linewidth]{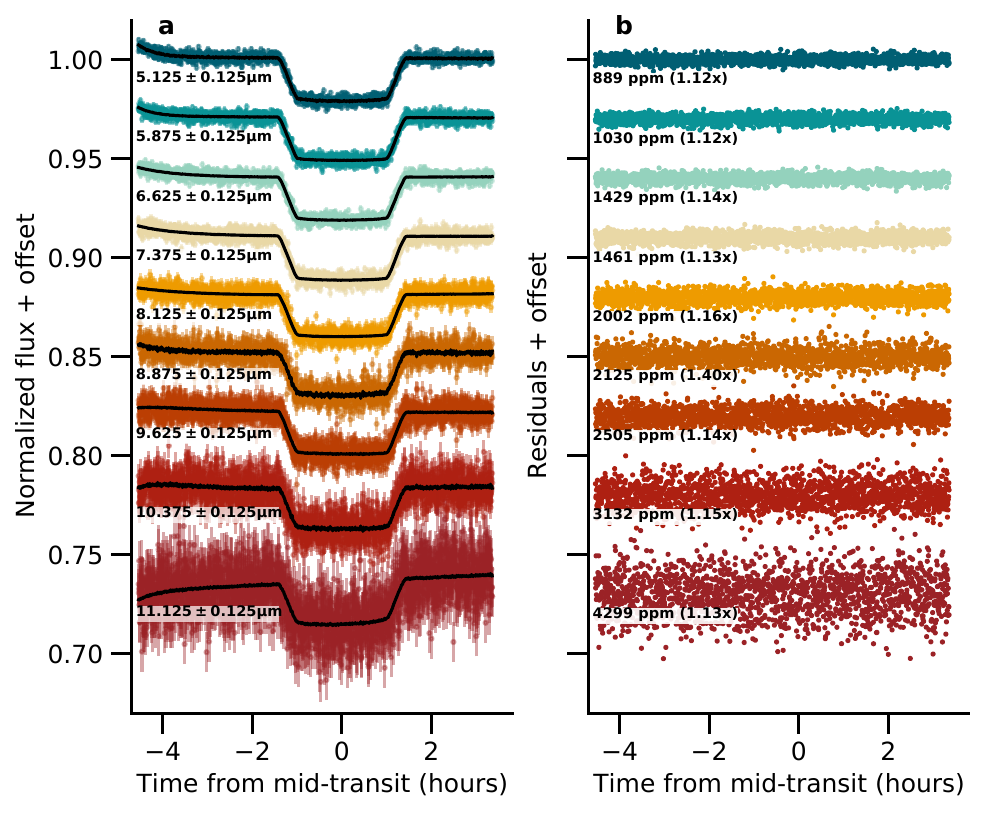}
\end{figure}
\noindent\textbf{Fig. 1 A sample of spectrophotometric light curves and residuals for WASP-39b's transit observed with MIRI/LRS.} \textbf{a:} An exoplanet transit model multiplied by a systematics model (solid black line) was fitted to each light curve. \textbf{b:} The residuals to the best-fit models are shown for each light curve. We report the $1\sigma$ scatter in each light curve as the standard deviation of the out-of-transit residuals, with the ratio to the predicted photon noise in parentheses. The reduction is from \texttt{Eureka!}.

\hfill \pagebreak

\begin{figure}[hbt!]
\centering
\includegraphics[width=0.99\linewidth]{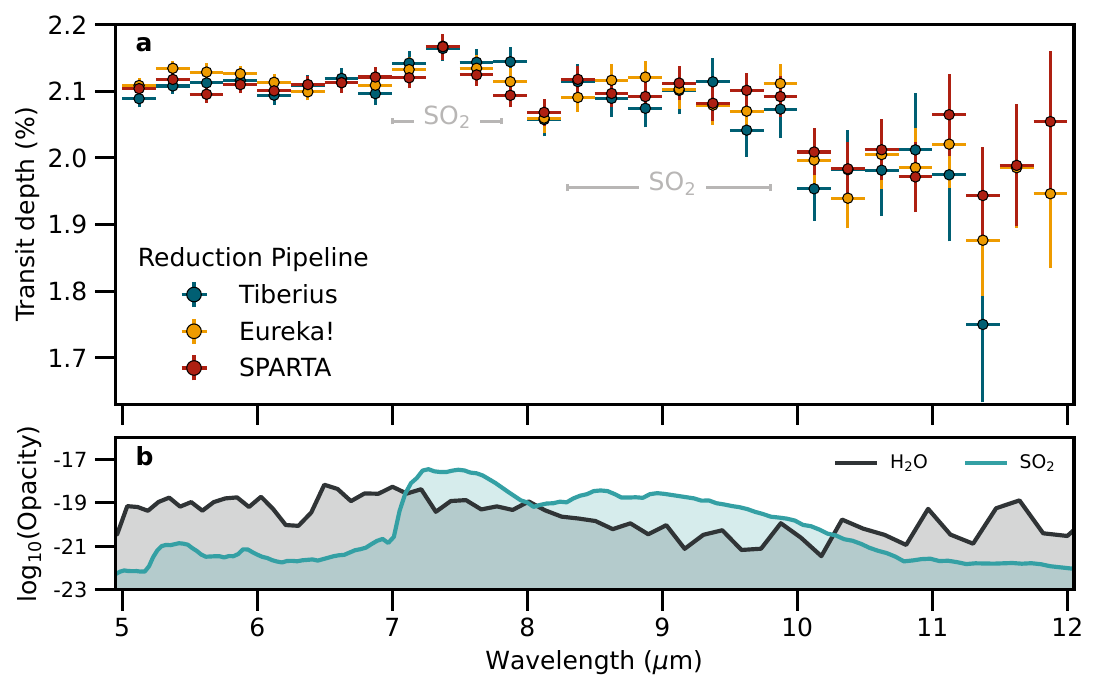}
\end{figure}
\noindent\textbf{Fig. 2 MIRI/LRS transmission spectra of WASP-39b derived using three independent reduction pipelines. a:} The spectrum is dominated by broad absorption features from SO$_2$ at 7.7 and 8.5~$\mu$m and H$_2$O across the entire wavelength coverage of MIRI/LRS. We define our uncertainties as $1\sigma$. \textbf{b:} We present the log of opacities of dominant species in the spectrum in units of cm$^2$ mol$^{-1}$. The opacities were adopted from \texttt{PLATON} using ExoMol line lists \citep{Polyansky2018,Underwood2016} and assume atmospheric properties pressure, $P = 1$\,mbar and temperature, $T = 1000$\,K.

\hfill \pagebreak

\begin{figure}[hbt!]
\centering
\includegraphics[width=0.99\linewidth]{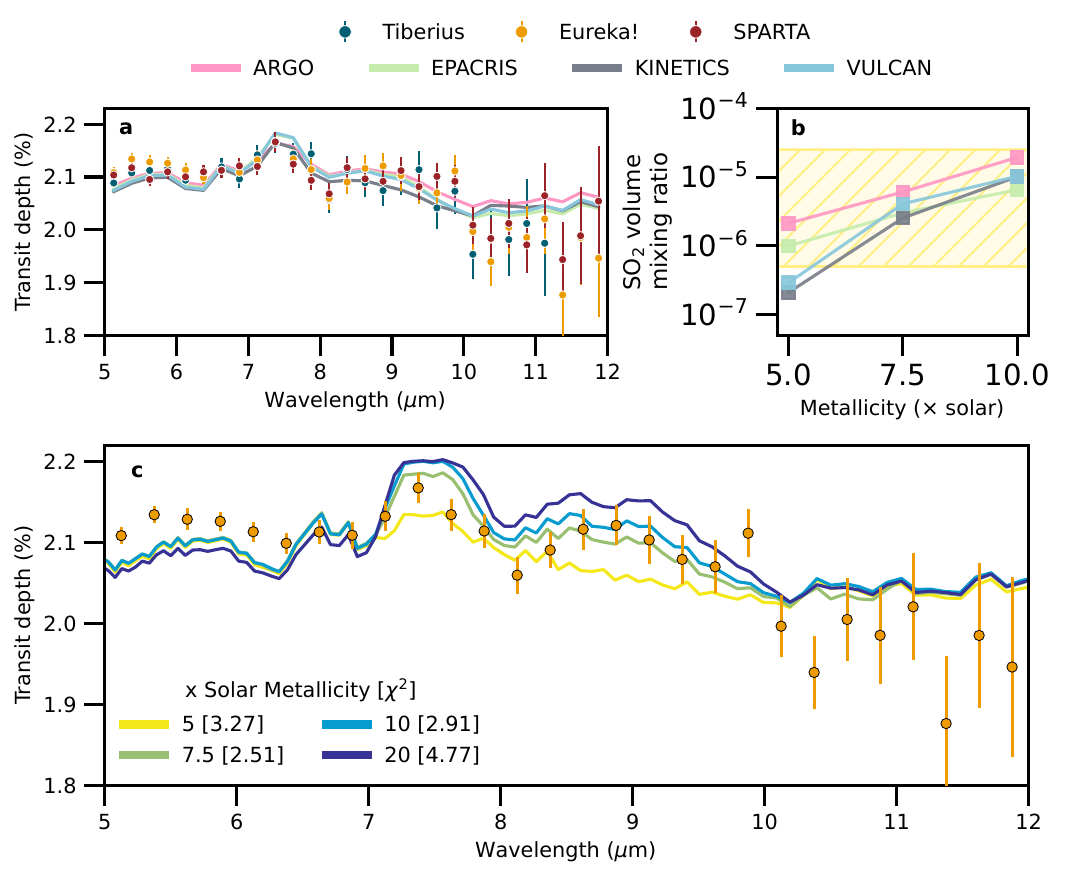}
\end{figure}
\noindent\textbf{Fig. 3 Free retrievals of the MIRI/LRS transmission spectrum of WASP-39b. a:} The spectrum from the \texttt{Eureka!} reduction (with 1$\sigma$ uncertainties) is compared to the best-fit retrieved spectra and associated 1$\sigma$ shaded regions from six free retrieval codes. \textbf{b:} The corresponding posterior probability distributions of the volume mixing ratio (VMR) and associated 1$\sigma$ uncertainties (points) for the \ce{SO2} abundance. The quoted log(\ce{SO2}) ranges from the lowest to the highest 1$\sigma$ bounds of all six posteriors. We chose the \texttt{Eureka!} reduction due to its similar reduction steps to previous WASP-39\,b observations \citep{ahrer2023,Alderson2023,feinstein2023,rustamkulov2023} and the fact that it provides the full wavelength coverage of the observations. Results from the other two reductions for SO$_2$ give broadly consistent results and are discussed further in Methods.

\hfill \pagebreak

\begin{figure}[hbt!]
\centering
\includegraphics[width=0.99\linewidth]{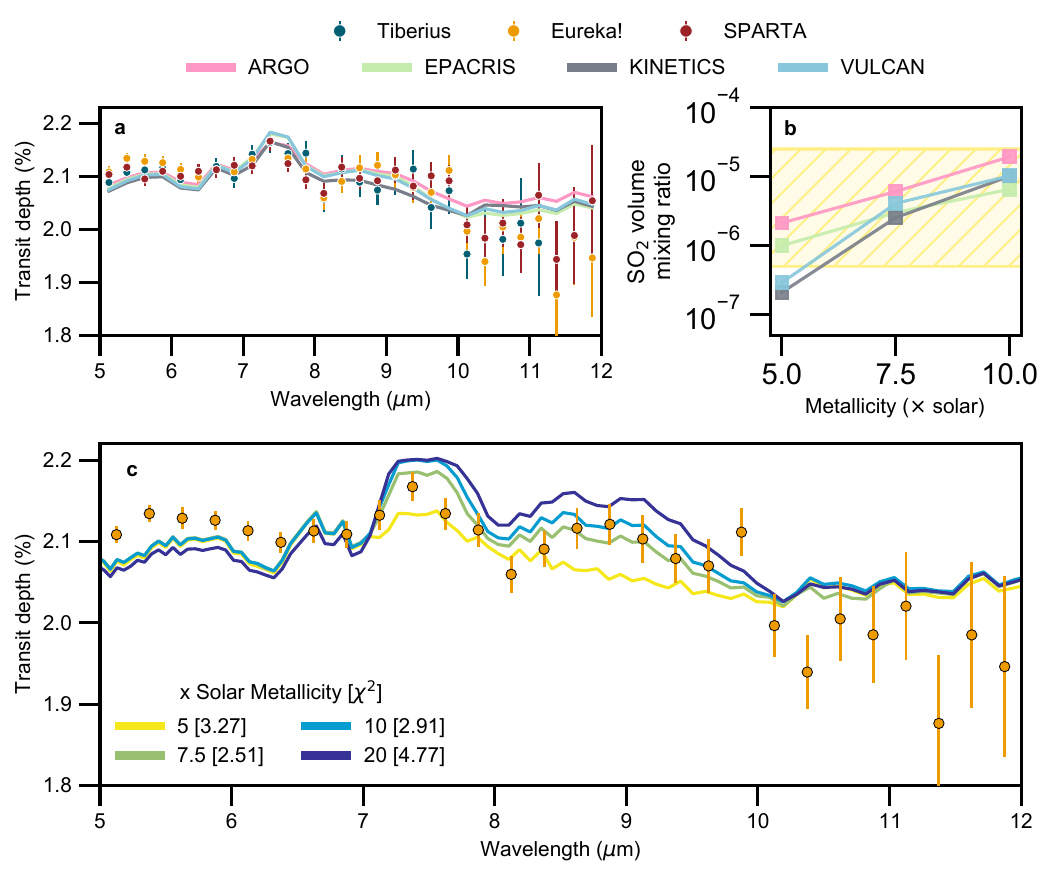}
\end{figure}
\noindent\textbf{Fig. 4 Comparison of four independent photochemical models to the observed MIRI/LRS transmission spectra of WASP-39b.} \textbf{a:} Comparison of morning and evening limb-averaged theoretical transmission spectra to the observations assuming a best-fit atmospheric metallicity of 7.5 $\times$ solar. \textbf{b:} Limb-averaged \ce{SO2} VMR between 10 and 0.01 mbar as a function of metallicity for the four photochemical models. The shaded and hatched yellow region represents the 1$\sigma$ \ce{SO2} constraint from the free retrievals on the \texttt{Eureka!} reduction (Fig. 3) . 
    \textbf{c:} Dependence of VULCAN modeled transmission spectrum on atmospheric metallicity, as compared to the \texttt{Eureka!} reduction. The Tiberius reduction prefers a metallicity of $7.5\times$ solar, while the SPARTA reduction prefers $10\times$ solar (see Extended Data). The VULCAN models suggest that there is only a minor ($< 0.05\%$) difference expected for the \ce{SO2} feature at $7.7\mu$m when assuming a higher atmospheric metallicity, while the \ce{SO2} feature at $8.5\mu$m is more sensitive to subtle changes. The \ce{SO2} feature at $8.5 \mu$m is fit well by the $7.5 - 10 \times$ solar metallicity models. 

\hfill \pagebreak

\begin{figure}[hbt!]
\centering
\includegraphics[width=0.99\linewidth]{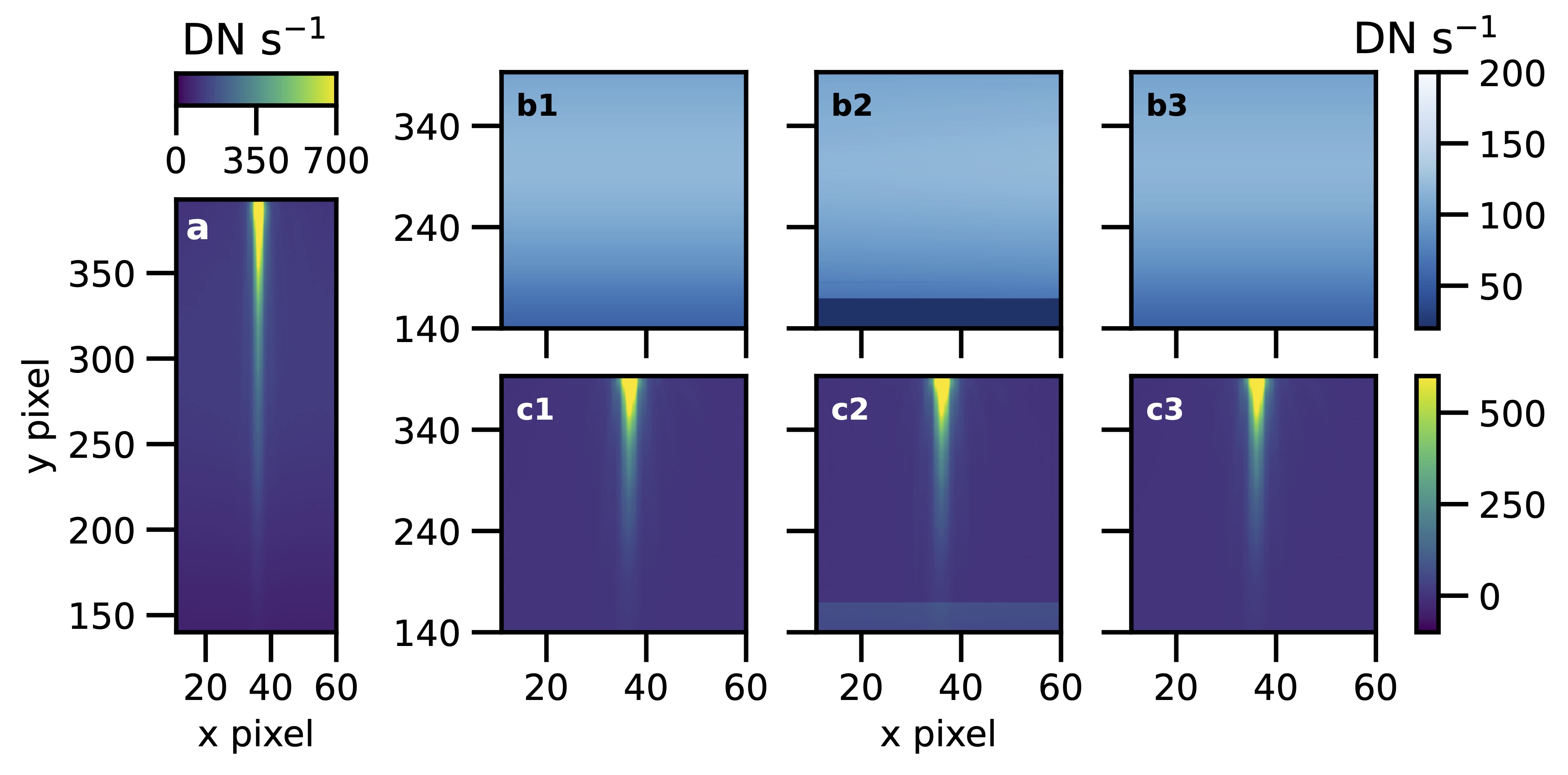}
\end{figure}
\noindent\textbf{Extended Data Fig. 1 Comparison of the different background modelling and subtraction per each pipeline.} (a) A median out-of-transit image of the MIRI/LRS detector from the \texttt{jwst} pipeline's Stage 2 processing. (b) Background models from \texttt{Eureka!} (1), Tiberius (2), and SPARTA (3). (c) Background subtracted Stage 2 outputs from each pipeline. The smoothly varying background is expected for MIRI/LRS. There are no discrete features or sharp changes in the background at y-pixels $< 244$, corresponding to $\lambda = 10$\,$\mu$m, which has been seen in other observations \citep{bell2023_arxiv}. All images are given in Data Numbers per second (DN s$^{-1}$). The Tiberius reduction did not extract spectra as far red as \texttt{Eureka!} and SPARTA, which is the cause of the horizontal bar in panels b2 and c2. 

\hfill \pagebreak

\begin{figure}[hbt!]
\centering
\includegraphics[width=0.99\linewidth]{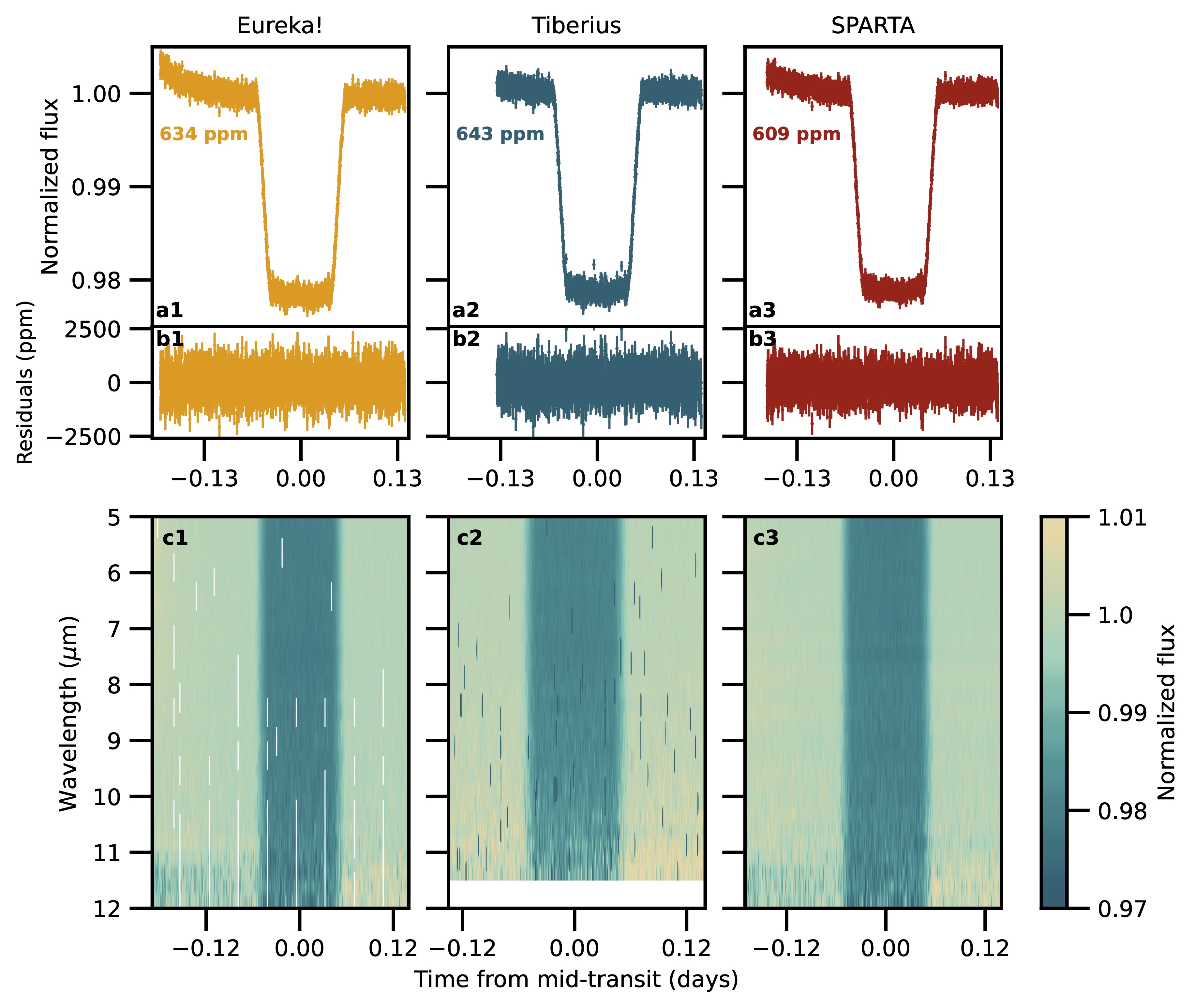}
\end{figure}
\noindent\textbf{Extended Data Fig. 2 MIRI/LRS white and spectrophotometric light curves from the three independent reduction pipelines used in this work.} (a) We quote the out-of-transit parts-per-million scatter in each light curve in the figure. We define the out-of-transit time as $-0.135 < t \textrm{ [days]} < -0.07$ and $0.07 < t \textrm{ [days]} < 0.14$; these times were selected as they ignore the exponential ramp at the beginning of the observations and do not include any data in transit ingress/egress. (b) The residuals and errors of the data compared to the best-fit transit model. Errors quoted are $1\sigma$. (c) The spectrophotometric light curves are normalized by the out-of-transit flux during the observations. All reductions show consistent out-of-transit scatter in all wavelength bins ($\Delta \lambda = 0.25$\,$\mu$m). The white spaces in c1 are where values in the light curve are NaN.

\hfill \pagebreak

\begin{figure}[hbt!]
\centering
\includegraphics[width=0.99\linewidth]{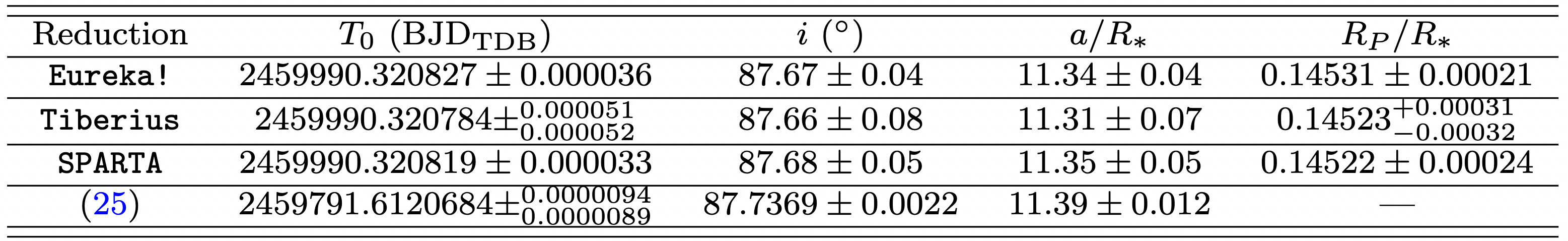}
\end{figure}
\noindent\textbf{Extended Data Table 1} The system parameters resulting from the white light curve fits.

\hfill \break
\begin{figure}[hbt!]
\centering
\includegraphics[width=0.99\linewidth]{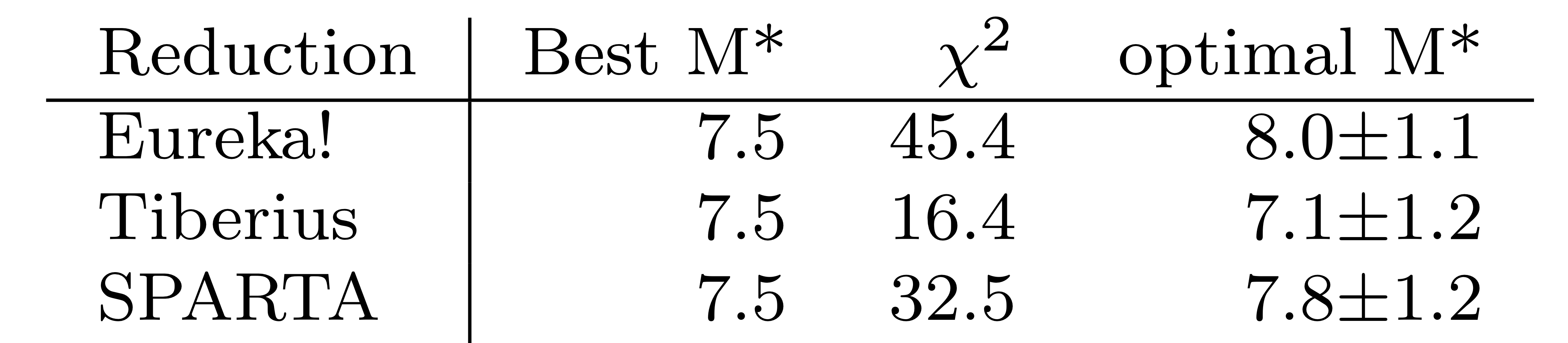}
\end{figure}

\noindent\textbf{Extended Data Table 2} Results from the IDIC grid assuming C, O, and S have the same abundance enhancement relative to Solar (i.e., M*).

\hfill \break

\begin{figure}[hbt!]
\centering
\includegraphics[width=0.99\linewidth]{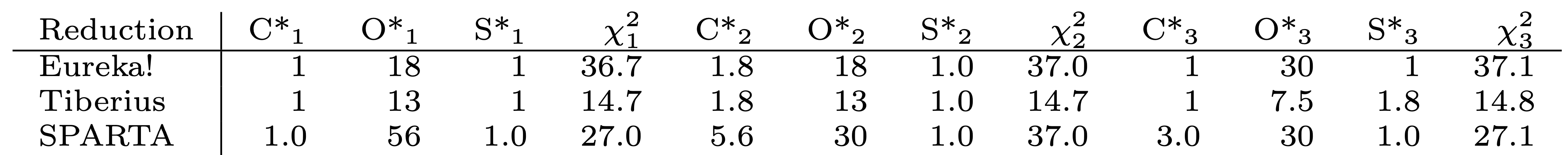}
\end{figure}
\noindent\textbf{Extended Data Table 3} Results from the IDIC grid assuming C, O, and S can take different abundances relative to Solar (C*, O*, S*). $\chi^2$ for the three best-fitting model spectra for each of the three reductions are shown.

\hfill \pagebreak

\begin{figure}[hbt!]
\centering
\includegraphics[width=0.99\linewidth]{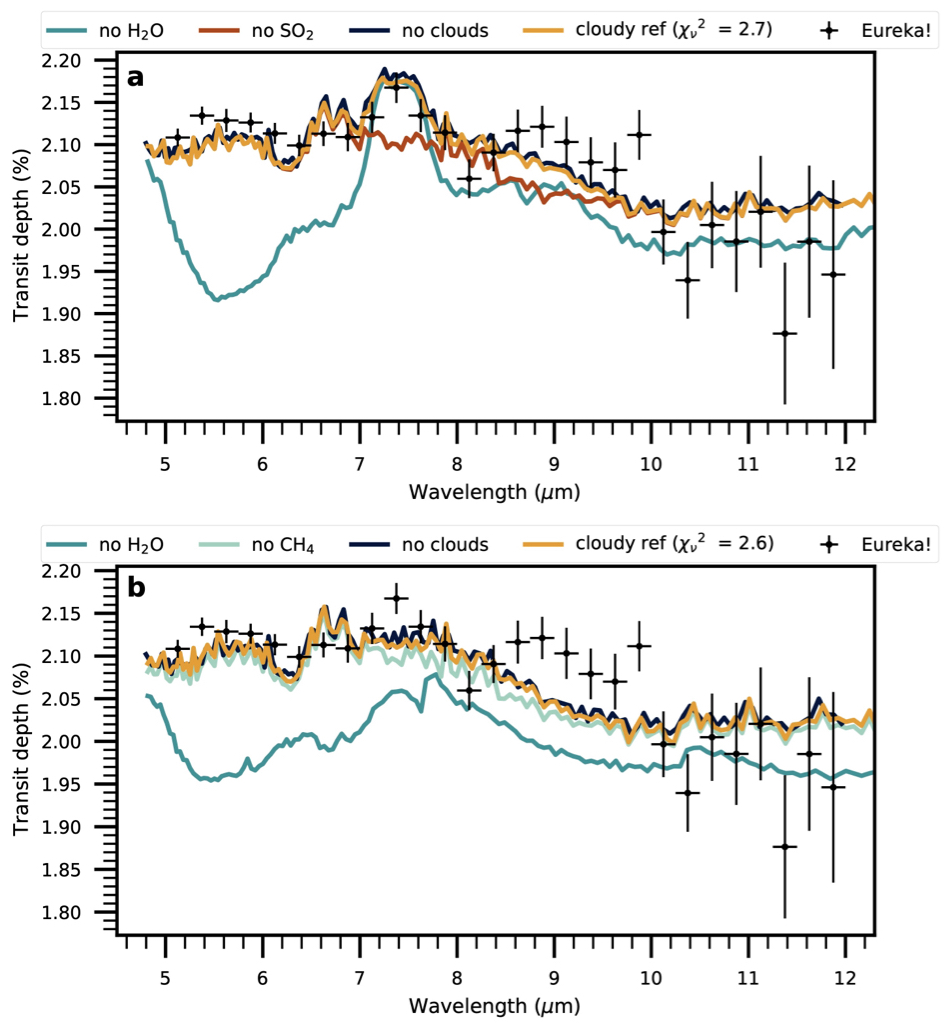}
\end{figure}
\noindent\textbf{Extended Data Fig. 3 The best-fitting cloudy \texttt{PICASO} grid models (gold lines) are shown with SO$_2$ (a) and without SO$_2$ (b) compared to the JWST MIRI/LRS data (black points) from the \texttt{Eureka!} reduction.} Also shown are the best-fits with \ce{H2O} (dark teal), \ce{SO2} (red), \ce{CH4} (light teal), and clouds (navy blue) removed from the model, demonstrating which absorbers dominate the opacity of the best-fit model. When SO$_2$ is not included in the model, excess \ce{CH4} compensates for its absorption in the \texttt{Eureka!} reduction as shown in the lower panel.

\hfill \pagebreak

\begin{figure}[hbt!]
\centering
\includegraphics[width=0.8\linewidth]{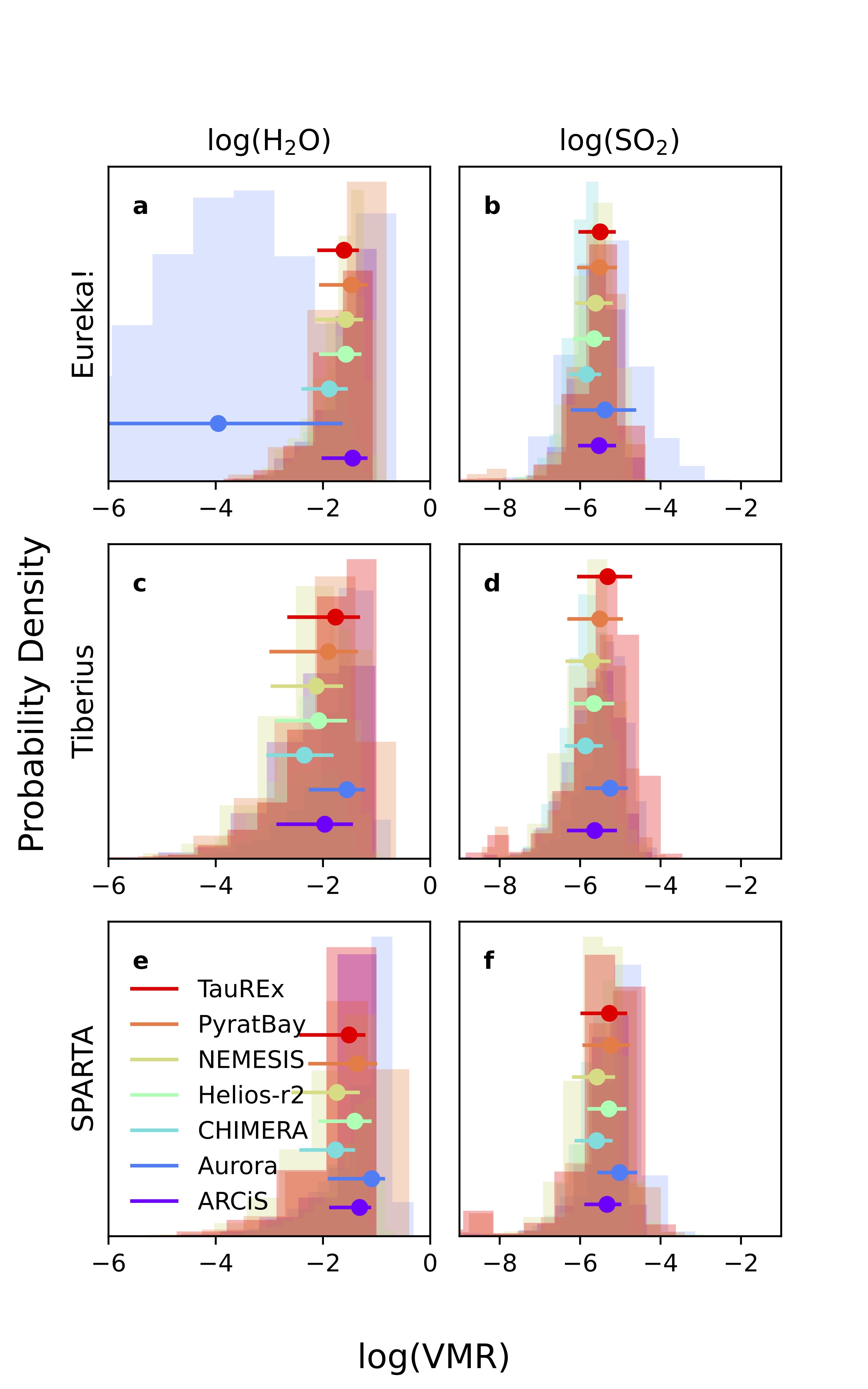}
\end{figure}
\noindent\textbf{Extended Data Fig. 4 Retrieved log of \ce{SO2} and \ce{H2O} volume mixing ratio (VMR) posteriors from all six retrieval codes and three data reductions}. Median values and 1$\sigma$ uncertainties are given in the coloured points. 

\hfill \pagebreak

\begin{figure}[hbt!]
\centering
\includegraphics[width=0.99\linewidth]{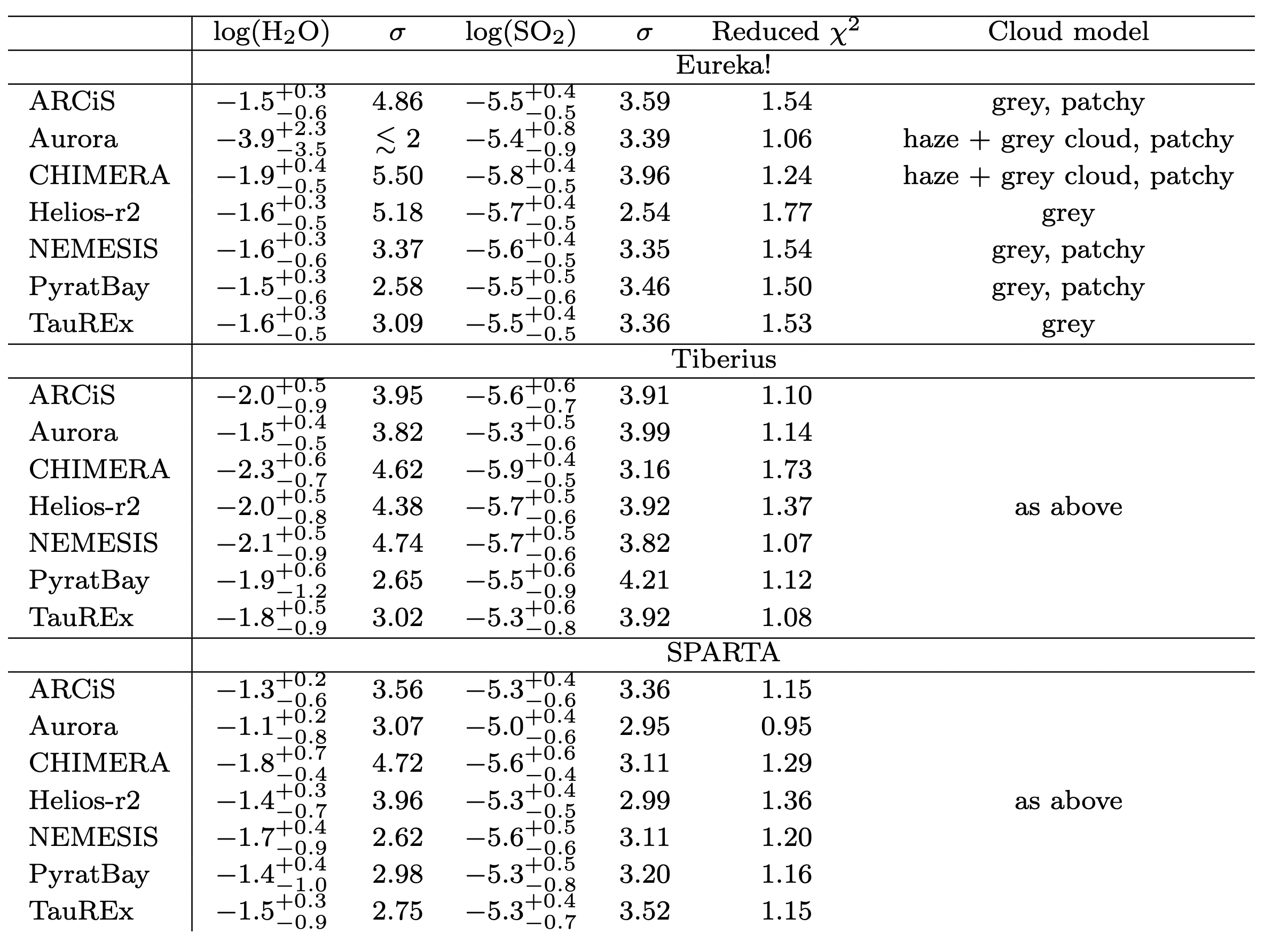}
\end{figure}
\noindent\textbf{Extended Data Table 4} This table collects all the free retrieval results for \ce{H2O} and \ce{SO2} volume mixing ratios, together with their detection significance, and the goodness of fit for each individual retrieval. The cloud model used for each retrieval code is also noted. For the most part, the abundances are consistent between retrieval codes for a given reduction, although there is some variation between reductions.

\newpage
\backmatter
\section*{Methods} \label{sec:methods}
\setcounter{page}{1}
\setcounter{figure}{0}
\renewcommand{\figurename}{Extended Data Fig.}
\renewcommand{\tablename}{Extended Data Table}

\subsection*{Data Reduction}

We applied three independent data reduction and light-curve-fitting routines to the MIRI/LRS observations. Below, we describe the major reduction steps taken by each pipeline, followed by their light-curve-fitting methodologies. Additionally, we discuss the differences in the data reduction pipelines that resulted in differing shapes of the \ce{H2O} absorption feature at $< 7 \mu$m.

\subsubsection*{\texttt{Eureka!}}

Initially, nine independent teams performed a reduction of these data using the open-source \texttt{Eureka!}\citeApp{bell2022} pipeline. From those analyses, we ultimately chose one analysis to highlight in this paper based on comparisons of the white and red noise of the residuals after fitting. Our fiducial \texttt{Eureka!} reduction very closely followed the methods developed for the Transiting Exoplanet ERS team's MIRI/LRS phase curve observations of WASP-43b and described in ref.\citeApp{bell2023_arxiv, bell2023_Nature}. As extensive parameter studies were performed on \texttt{Eureka!}'s Stage 1--3 parameters using the WASP-43b data, the best parameter settings identified from that work are reused here and are briefly summarized below. The other \texttt{Eureka!} analyses had used different reduction parameters and were generally consistent with, but noisier than, our fiducial \texttt{Eureka!} analyses. The full \texttt{Eureka!} Control Files and \texttt{Eureka!} Parameter Files files used in these analyses are available as part of the data products associated with this work (https://doi.org/10.5281/zenodo.10055845).

We made use of version 0.9 of the \texttt{Eureka!}\citeApp{bell2022} pipeline, CRDS version 11.16.16 and context 1045, and \texttt{jwst} package version 1.8.3 \citeApp{jwst_v1.8.2}. As described in ref.\ \citeApp{bell2023_arxiv, bell2023_Nature}, we assume a constant gain of 3.1 electrons/DN (same as for the SPARTA reduction; see below), which is closer to the true gain than the value of 5.5 currently assumed in the CRDS reference files (private comm., Sarah Kendrew). \texttt{Eureka!}'s Stage 1 jump step's rejection threshold was increased to 7.0 and Stage 2's photom step was skipped (to more easily estimate the expected photon noise), but otherwise the Stage 1--2 processing was done following the \texttt{jwst} pipeline's default settings. We also evaluated the use of an experimental non-linearity reference file developed to address MIRI's ``brighter-fatter effect''\citeApp{argyriou_2023}, but we ultimately decided to stick with the default non-linearity reference file as the final transmission spectra changed by less than $1\sigma$ at all wavelengths.

We extracted columns 11--61 and rows 140--393 as pixels outside of this range are excessively dominated by noise. We masked pixels marked as ``DO\textunderscore NOT\textunderscore USE'' in the DQ array to remove bad pixels identified by the \texttt{jwst} pipeline. To aid in decorrelating systematic noise, we compute a single centroid and PSF-width for each integration by summing along the dispersion direction and fitting a 1D Gaussian; only the first integration's centroid was used to determine aperture locations. We subtracted the background flux by subtracting the mean of pixels separated from the source by 11 or more pixels after first sigma-clipping $5\sigma$ outliers along the time axis and along the spatial axis. We then performed optimal spectral extraction \citeApp{horne1986optspec} using the pixels within 5 pixels of the centroid. Our spatial profile was a cleaned median frame, following the same sigma-clipping methods described by ref.\ \citeApp{bell2023_arxiv, bell2023_Nature}. We then spectrally binned the data into 28 bins, each 0.25~$\mu$m wide, spanning 5--12~$\mu$m as well as a single white light curve spanning the full 5--12~$\mu$m. To remove any remaining cosmic rays or the effects of any high-gain antenna moves, we then sigma-clipped each light curve, removing any points 4 or more sigma discrepant with a smoothed version of the light curve computed using a box-car filter with a width of 20 integrations. This removed errant points while ensuring not to clip the transit ingress or egress.

When fitting, our astrophysical model consisted of a \texttt{starry} \citeApp{starry} transit model with uninformative priors on the planet-to-star radius ratio and unconstrained, reparameterized quadratic limb-darkening parameters \citeApp{Kipping2013}. We also used broad priors on the planet's orbital parameters to verify that these new data are consistent with the orbital solution presented by ref.\citeApp{datasynthesis2023}. Specifically, we used Gaussian priors for the transit time, inclination, and scaled semi-major axis based on the values of ref.\citeApp{datasynthesis2023} which were derived by fitting all previous WASP-39b observational datasets at once, see values in Extended Data Table 1, but with greatly inflated uncertainties (roughly 10$\times$ or higher than the precision achievable with these MIRI data alone) to allow these data to independently verify the previously published values \citeApp{datasynthesis2023}. We also assumed zero eccentricity and fixed the orbital period to the value of 4.0552842 $\pm^{0.0}_{0.0000035}$ days from ref.\citeApp{datasynthesis2023}. We linearly decorrelated against the changing spatial position and PSF-width computed during Stage 3. We also allowed for a linear trend in time as well as a single weakly constrained exponential ramp to remove the well-known ramp at the beginning of MIRI/LRS observations \citeApp{bouwman2023, bell2023_arxiv, bell2023_Nature}. We also trimmed the first 10 integrations as they suffered from a particularly strong exponential ramp. There was no evidence for mirror tilts \citeApp{schlawin2023} in the observations nor any residual impacts from high-gain antenna moves after sigma-clipping the data in Stage 4. Finally, we also used a noise multiplier to capture any excess white noise and ensure a reduced chi-squared of 1. We then used \texttt{PyMC3}'s No U-Turns Sampler \citeApp{pymc3} to sample our posterior. We used two independent chains and used the Gelman--Rubin statistic \citeApp{GelmanRubin1992} to ensure that our chains had converged ($\hat{R}<1.01$), and then we combined the samples from the two chains and computed the 16th, 50th, and 84th percentiles of the 1D marginal posteriors to estimate the best-fit value and uncertainty for each parameter.

As our determined orbital parameters were consistent with those determined by ref.\citeApp{datasynthesis2023}, we then fixed our orbital parameters to those of ref.\citeApp{datasynthesis2023} for our spectroscopic fits ensuring consistency with other JWST spectra for this planet. The limb-darkening parameters for our spectroscopic fits were given a Gaussian prior of $\pm$0.1 with respect to model-predicted limb-darkening coefficient spectra \citeApp{Morello2020_aj, Morello2020_joss} based on the Stagger grid \citeApp{Chiavassa2018}. We also evaluated more conservatively trimming the first 120 integrations (instead of 10) for our spectroscopic fits, but found that the resulting spectra were changed by much less than $1\sigma$ at all wavelengths.

For our white light curve fit, we found a white noise level 26\% larger than the estimated photon limit, while the spectroscopic channels were typically 10--20\% larger than the estimated photon limit. As our adopted gain of 3.1 is only accurate to within $\sim$10\% of the true gain (which varies as a function of wavelength; private comm., Sarah Kendrew; \citeApp{bell2023_arxiv, bell2023_Nature}), these comparisons to estimated photon limits only give general ideas of MIRI’s performance. An examination of our Allan variance plots \citeApp{Allan1966} showed minimal red noise in our residuals. Our decorrelation against the spatial position and PSF-width showed that the shortest wavelengths were most strongly affected by changes in spatial position and PSF-width, with both driving noise at the level of $\sim$100\,ppm in the shortest wavelength bin; meanwhile, the impact at longer wavelengths was weaker and not as well constrained. The orbital parameters determined from the white light curve fit are summarized in Extended Data Table 1.

\subsubsection*{Tiberius}

\texttt{Tiberius} is a pipeline to perform spectral extraction and light-curve fitting, which is derived from the LRG-BEASTS pipeline \citeApp{Kirk2017, 2019AJ....158..144K, 2021AJ....162...34K}. It has been used in the analysis of JWST data from the ERS Transiting Exoplanet Community program and GO programs \citeApp{jwst2022,Alderson2023,rustamkulov2023,Lustig-Yaeger2023}.

In our reduction with \texttt{Tiberius}, we first ran STScI's \texttt{jwst} pipeline on the \texttt{uncal.fits} files. We performed the following steps in the \texttt{jwst} pipeline: \texttt{group\_scale}, \texttt{dq\_init}, \texttt{saturation}, \texttt{reset}, \texttt{linearity}, \texttt{dark\_current}, \texttt{refpix}, \texttt{ramp\_fit}, \texttt{gain\_scale}, \texttt{assign\_wcs} and \texttt{extract\_2d}. Our spectral extraction was run on the \texttt{gainscalestep.fits} files and we used the \texttt{extract2d.fits} files for our wavelength calibration. As explained in the \texttt{jwst} documentation, the \texttt{gain\_scale} step is actually benign if the default gain setting is used. For that reason, the \texttt{Tiberius} reduction used units of DN/s. Ultimately, since we normalize our light curves and rescale the photometric uncertainties during light curve fitting, the units of the extracted stellar flux do not impact the transmission spectrum. 

We did not perform the \texttt{jump} or \texttt{flat\_field} steps. Instead of the jump step, we performed outlier detection for every pixel in the time-series by locating integrations for which a pixel deviated by $> 5\sigma$ from the median value for that pixel. Any outlying pixels in the time-series were replaced by the median value for that pixel. Next we performed spectral extraction. We first interpolated the spatial dimension of the data onto a new grid with $10\times$ the resolution, which improves flux extraction at the sub-pixel level. The spectra were then traced using Gaussians fitted to every pixel row from row 171 to 394. The means of these Gaussians were then fitted with a fourth-order polynomial. We then performed standard aperture photometry at every pixel row after subtracting a linear polynomial fitted across two background regions on either side of the spectral trace. We experimented with the choice of aperture width and background width to minimize the noise in the white light curve. The result was a 8-pixel-wide aperture and two 10-pixel-wide background regions offset by 8 pixels from the extraction aperture. 

Next we cross-correlated each integration's stellar spectrum with a reference spectrum to measure drifts in the dispersion direction. The reference spectrum was taken to be the 301st integration of the time-series, as we clipped the first 300 integrations (80 minutes) to remove the ramp seen in the transit light curve. The measured shifts had an RMS of 0.002 pixels in the dispersion direction and 0.036 in the spatial direction (as measured from the tracing step). Next we integrated our spectra in $25\times$ 0.25\,$\mu$m-wide bins from 5--11.25\,$\mu$m to make our spectroscopic light curves. 

We fitted our light curves with an analytic transit light curve, implemented in \texttt{batman} \citeApp{batman}, multiplied by a time trend. For the white light curve, this time trend was a quadratic polynomial, as a linear trend was not sufficient. This differed to the other reductions that treated the systematics as exponential ramps with a linear trend. For the spectroscopic light curves, we divided each spectroscopic light curve by the best-fitting transit and systematics model from the white light curve fit. A quadratic trend was not necessary for the spectroscopic light curves, which we instead fit with a linear trend to account for residual chromatic trends not accounted for by the common mode correction.

In all light curve fits, we used Markov Chain Monte Carlo (MCMC) implemented via \texttt{emcee} \citeApp{emcee2013}. We set the number of walkers equal to $10\times$ the number of free parameters and ran two sets of chains. The first set of chains was used to rescale the photometric uncertainties to give $\chi^2_{\nu} = 1$ and the second set of chains was run with the rescaled uncertainties. In both cases, the chains were run until they were at least $50\times$ the autocorrelation length for each parameter. This led to chains between 4000--10000 steps long.  

Given the non-linear ramp at the beginning of the observations, we clipped the first 300 integrations. We found this clipping led to a consistent and more precise transmission spectrum. In tests without clipping any integrations, we found that a fifth order polynomial was needed to fit the ramp. We disfavoured this due to the extra free parameters. For the white light curve, our fitted parameters were the time of mid-transit ($T_0$), orbital inclination of the planet ($i$), semi-major axis scaled by the stellar radius ($a/R_*$), planet-to-star radius ratio ($R_P/R_*$), the three parameters defining the quadratic-in-time polynomial trend, and the quadratic limb darkening coefficients reparameterized following \citeApp{Kipping2013} ($q1$ and $q2$). For $q1$ and $q2$ we used Gaussian priors with means set by calculations from Stagger 3D stellar atmosphere models \citeApp{Chiavassa2018,Morello2020_aj, Morello2020_joss} and standard deviations of 0.1. The period was fixed to 4.0552842518\,d as found from the global fit to the near-IR JWST datasets \citeApp{datasynthesis2023}. Our best-fitting values for the system parameters are given in Table 1. 

For our spectroscopic light curves, we fixed the system parameters ($a/R_*$, $i$, $T_0$) to the values from the global fit to the near-IR JWST datasets \citeApp{datasynthesis2023}. The median RMS of the residuals from the white light and spectroscopic light curve fits were 573 and 3034\,ppm, respectively.

\subsubsection*{SPARTA}
The Simple Planetary Atmosphere Reduction Tool for Anyone (SPARTA) is an open-source code intended to be simple, fast, barebones, and utilitarian.  SPARTA is fully independent and uses no code from the JWST pipeline or any other pipeline.  It was initially written to reduce the MIRI phase curve of GJ\,1214b, and is described in detail in that paper \citeApp{kempton_2023}.  SPARTA was also used to reduce the MIRI phase curve of WASP-43b, taken as part of the Early Release Science program \citeApp{bell2023_arxiv, bell2023_Nature}.  Having learned many best practices from these previous reductions, we performed virtually no parameter optimization for the current WASP-39b reduction.  Below, we briefly summarize the reduction steps, but we refer the reader to the previous two papers for more details.

In stage 1, SPARTA starts with the uncalibrated files and performs nonlinearity correction, dark subtraction, up-the-ramp fitting, and flat correction, in that order.  The up-the-ramp fit discards the first 5 groups and the last group, which are known to be anomalous, and optimally estimates the slope using the remaining groups by taking the differences between adjacent reads and computing the weighted average of the differences.  The weights are calculated with a mathematical formula which gives the optimal estimate of the slope \citeApp{kempton_2023}.

After stage 1, SPARTA computes the background by taking the average of columns 10--24 and 47--61 (inclusive, zero-indexed) of each row in each integration.  The background is then subtracted from the data.  These two windows are equally sized and equidistant from the trace on either side, so any slope in the background is naturally subtracted out.

Next, we compute the position of the trace.  We compute a template by taking the pixel-wise median of all integrations.  For each integration, we shift the template (via bilinear interpolation) and scale the template (via multiplication by a scalar) until it matches the integration.  The shifts that result in the lowest $\chi^2$ are recorded.

The aforementioned template, along with the positions we find, are used for optimal extraction.  We divide the template by the per-row sum (an estimate of the spectrum) to obtain a profile, and shift the profile in the spatial direction by the amount found in the previous step.  The shifted profile is then used for optimal extraction, using the algorithm of \citeApp{horne1986optspec}.  We apply this algorithm only to a 11-pixel-wide (full width) window centered on the trace, and iteratively reject $>5\sigma$ outliers until convergence.

After optimal extraction, we gather all the spectra and the positions into one file.  We reject outliers by creating a white light curve, detrending it with a median filter, and rejecting integrations $>4\sigma$ away from 0.  Sometimes, only certain wavelengths of an integration are bad, not the entire integration.  We handle these by detrending the light curve at each wavelength, identifying 4$\sigma$ outliers, and replacing them with the average of their neighbors on the time axis.

Finally, we fit the white light and spectroscopic light curves using \texttt{emcee}.  The spectroscopic bins are exactly the same as for the Eureka! and Tiberius reductions: 0.25 $\mu m$ wide and ranging from 5.00--5.25 $\mu m$ to 11.75--12.00 $\mu m$.  We trim the first 112 integrations (30 minutes), and reject $>4\sigma$ outliers.  In the white light fit, limb darkening parameters $q_1$ and $q_2$ are both free and given broad uniform priors.  In the spectroscopic fit, $T_0$, $P$, $a/R_s$, $b$, and the limb darkening coefficients are fixed to the fiducial values, but the transit depth and the systematics parameters are free.  The systematics model is given by
\begin{align}
    S = F_* (1 + A\exp{(-t/\tau)} + c_y y + c_x x + m(t-\overline{t})),
\end{align}
where $F_*$ is a normalization constant, A and $\tau$ parameterize the exponential ramp, t is the time since the beginning of the observations (after trimming), x and y are the positions of the trace on the detector, m is a slope (potentially caused by stellar variability and/or instrumental drift), and $\overline{t}$ is the average time. All parameters are given uniform priors.  $\tau$ is required to be between 0 and 0.1, but no explicit bounds are imposed on the other parameters.

\subsection*{Forward Modelling}
We used several forward models that take into account photochemistry to infer the properties of WASP-39b's atmosphere from the observations. These models are based on known first-principle physics and chemistry that aid in our understanding of the important atmospheric processes at work. In addition, we also use one of the models to generate a more extensive model grid to assess the atmospheric metallicity and elemental ratios of WASP-39b. These models compute the atmospheric composition by explicitly treating the thermochemical and photochemical reactions and transport in the atmosphere, and in general are initialized from equilibrium abundances based on a given elemental ratio, for which we scale relative to Solar abundances \citeApp{Lodders2020}. Although the abundances of a planet's host star are the more natural comparison point \citeApp[e.g.,][]{pacetti:2022}, the measured multi-element abundances of WASP-39 are very nearly Solar \citeApp{Polanski2022}. All photochemical models use the same incident stellar spectrum as that described in ref.\ \citeApp{Tsai2023}. Finally, we also consider a radiative-convective thermochemical equilibrium model that includes an injected \ce{SO2} abundance and clouds to connect our work to previous interpretations of near-infrared JWST spectra of WASP-39b \citeApp{rustamkulov2023,feinstein2023,ahrer2023,Alderson2023}. 

\subsubsection*{VULCAN}
The 1D kinetics model VULCAN treats thermochemical \citeApp{tsai17} and photochemical \citeApp{Tsai2021} reactions. VULCAN solves the Eulerian continuity equations including chemical sources/sinks, diffusion and advection transport, and condensation. We used the C–H–N–O–S network (\url{https://github.com/exoclime/VULCAN/blob/master/thermo/SNCHO_photo_network.txt}) for reduced atmospheres containing 89 neutral C-, H-, O-, N-, and S-bearing species and 1028 total thermochemical reactions (i.e., 514 forward-backward pairs) and 60 photolysis reactions. The sulphur allotropes are simplified into a system of S, \ce{S2}, \ce{S3}, \ce{S4}, and \ce{S8}. The sulphur kinetics data is drawn from the NIST and KIDA databases, as well as modelling \citeApp{Moses1996,Zahnle2016} and ab-initio calculations published in the literature \citeApp[e.g.,][]{Du2008}. The temperature-dependent UV cross sections \citeApp{Tsai2021} are not used in this work for simplicity, but preliminary tests show that their exclusion has resulted in only minor differences (less than 50\% of the \ce{SO2} VMR). Apart from varying elemental abundances, we applied an identical setup of VULCAN as that in ref.\ \citeApp{Tsai2023}.

\subsubsection*{KINETICS}
The \texttt{KINETICS} 1D thermo-photochemical transport model \citeApp{allen81,yung84,Moses11,Moses2013} is used to solve the coupled Eulerian continuity equations for the production, loss, and vertical diffusive transport of atmospheric species. The chemical reaction list, background atmospheric structure, and assumed planetary parameters are identical to those described in ref.\ \citeApp{Tsai2023}, except here we explore additional atmospheric metallicities. Briefly, the C-H-N-O-S-Cl network used for the WASP-39b \texttt{KINETICS} model contains 150 neutral species that interact with each other through 2350 total reactions, with the non-photolysis reactions being reversed through the thermodynamic principle of microscopic reversibility \citeApp{visscher11}.  

\subsubsection*{ARGO}
The 1D thermochemical and photochemical kinetics code, ARGO, originally utilised the Stand2019 network for neutral hydrogen, carbon, nitrogen and oxygen chemistry \citeApp{Rimmer2016,Rimmer2019}. ARGO solves the coupled 1D continuity equation including thermochemical-photochemical reactions and vertical transport. The Stand2019 network was expanded by ref.\ \citeApp{Rimmer2021} by updating several reactions, incorporating the sulphur network developed by ref.\ \citeApp{Hobbs2021}, and supplementing it with reactions from ref.\ \citeApp{Krasnopolsky2007} and ref.\ \citeApp{Zhang2012}, to produce the Stand2020 network. The Stand2020 network includes 2901 reversible reactions and 537 irreversible reactions, involving 480 species composed of H, C, N, O, S, Cl and other elements. 

\subsubsection*{EPACRIS}
EPACRIS (ExoPlanet Atmospheric Chemistry \& Radiative Interaction Simulator) is a general-purpose one-dimensional atmospheric simulator for exoplanets. EPACRIS has a root of the atmospheric chemistry model developed by Renyu Hu and Sara Seager at MIT \citeApp{hu2012,hu2013,hu2014}, and since then has been reprogrammed and upgraded substantially \citeApp[][and also Yang \& Hu 2023, in prep. mainly focusing on the validation of reaction rate-coefficients]{hu2019information, hu2021photochemistry}. We use the atmospheric chemistry module of EPACRIS to compute the steady-state chemical composition of WASP-39~b's atmosphere controlled by thermochemical equilibrium, vertical transport, and photochemical processes. The chemical network applied in this study includes 60 neutral  C-, H-, O-, and S-bearing species and 427 total reactions (i.e., 380 reversible reaction pairs and 47 photodissociation reactions). In this chemical model, \ce{SO2} volume mixing ratio is sensitive to two reactions which are (i) \ce{H2S} $\leftrightarrow$ \ce{HS} + \ce{H} and (ii) \ce{SO} + \ce{OH} $\leftrightarrow$ \ce{HOSO}). Briefly describing, if \ce{HS} + \ce{H} $\to$ \ce{H2S} recombination rate-coefficient is faster than $10^{-11}$ cm\textsuperscript{3}/molecule/s (collision-limit is around $10^{-9}$ cm\textsuperscript{3}/molecule/s), this will result in inefficient \ce{H2S} dissociation (i.e., \ce{H2S} starts to dissociate at higher altitude), which leads to the decreased \ce{SO2} formation. Unfortunately, to the best of our knowledge, there is no theoretically calculated nor experimentally measured \ce{H2S} decomposition rate coefficient. For this reason, in EPACRIS, we assumed that \ce{H2S} $\leftrightarrow$ \ce{HS} + \ce{S} is similar  to \ce{H2O} $\leftrightarrow$ \ce{HO} + \ce{H}. However, all the \ce{HS} + \ce{H} $\to$ \ce{H2S} recombination rate-coefficient used in different models were slower than $10^{-11}$ cm\textsuperscript{3}/molecule/s and below this range, \ce{SO2} volume mixing ratio isn't sensitive to this reaction anymore. With regard to the \ce{SO} + \ce{OH} $\leftrightarrow$ \ce{HOSO} reaction, the forward reaction (barrier-less reaction) is favored at lower temperatures and higher pressure according to the HOSO potential energy surfaces \citeApp{Hughes2002}. For this reason, the exclusion of this reaction from the EPACRIS chemical model shows up to 2 orders of magnitude increase (i.e., from [\ce{SO2}] $\sim$ $10^{-6}$ to $10^{-4}$) in the \ce{SO2} volume mixing ratio in the morning limb. However, in the evening limb whose temperature is up to $\sim$200 K higher compared to the morning limb, HOSO now can further dissociate to form \ce{SO2} and H due to elevated temperature, which results in the increased [\ce{SO2}] $\sim$ $10^{-5}$ compared to the morning limb [\ce{SO2}] $\sim$ $10^{-6}$.

\subsubsection*{IDIC Grid}

Ref.\ \citeApp{crossfield:2023} presented a grid of \texttt{VULCAN} photochemistry models (we term this the \href{https://memory-alpha.fandom.com/wiki/IDIC}{IDIC} grid) for WASP-39b that cover a 3D volume of possible C, O, and S elemental abundances without aerosols. We used these models to compare to our three spectral reductions.  We fit each MIRI/LRS transmission spectrum by binning all model spectra to the regular, 0.25\,$\mu$m resolution of the observed spectra,  allowing for an arbitrary vertical offset for each model spectrum, and calculating $\chi^2$ for each model spectrum. We first determined the goodness-of-fit while holding all abundances linked to the same value (i.e., C, O, and S all enhanced by the same level relative to Solar abundances). We fit a parabola to the three lowest $\chi^2$ points to estimate the optimal elemental abundance enhancement and its uncertainty \citeApp[i.e., where $\Delta \chi^2=1$;][]{avni:1976}. We then also compared these linked-abundance $\chi^2$ values to those derived across the entire 3D grid by allowing all three elemental abundances to vary individually. Extended Data Tables 2 and 3 show the abundances and $\chi^2$ values for these analyses.  

Interpreting the spectra is challenging because the goodness-of-fit varies widely across the observed spectra: across all IDIC models, we find a best-fit $\chi^2$ of 14.7 for the Tiberius reduction but a best-fit $\chi^2=45.4$ for the Eureka! reduction (which reports much smaller measurement uncertainties). Nonetheless the linked analyses all suggest a bulk metallicity of 7.1--8.0$\times$ Solar. The standard deviation of the optimal metallicity values is 0.4, smaller than the average uncertainties in Extended Data Table 2, suggesting that the uncertainty in the bulk metallicity is dominated by statistical (or model-dependent systematic) uncertainties rather than by differences between the several reduced spectra.  

When allowing C, O, and S abundances to each vary freely, in all cases the best-fitting models show a preference for super-solar O/S ratios, sub-solar C/O, and approximately solar C/S ratios.  Ref.~\citeApp{crossfield:2023} suggests that these ratios could be used to constrain a planet's formation history by comparing to formation models \citeApp{pacetti:2022,schneider:2021b}. However, a Bayesian Information Criterion (BIC) analysis shows that for the Tiberius and SPARTA reductions the observed spectra do not justify the additional free parameters of multiple independent elemental abundances. The formal BIC value for the Eureka! reduction seems to indicate that independent abundances are justified, but this conclusion seems questionable since this spectrum gives the worst $\chi^2$ values (36.7 with just 28 data points).

\subsubsection*{PICASO Grid}

Previous observations of WASP-39b with JWST's NIRspec PRISM, NIRISS SOSS, NIRCam F322W, and NIRSpec G395H \citeApp{jtec2023,ahrer2023,Alderson2023,feinstein2023,rustamkulov2023} were interpreted using a grid of 1D radiative-convective thermal equilibrium (RCTE) models \citeApp{Mukherjee2022Zenodo} generated with \texttt{PICASO 3.0} \citeApp{Batalha2019,Mukherjee2023}. Here, to interpret the spectrum of WASP 39b observed with MIRI LRS, we use the base clear equilibrium \texttt{PICASO 3.0} version of this grid along with a subset of the grid of \texttt{PICASO 3.0} models post-processed with \texttt{Virga} \citeApp{ackerman2001,Rooney2022} to account for clouds formed from Na$_2$S, MnS, and  MgSiO$_3$. The full parameters of the original set of grids can be found in ref.\ \citeApp{Mukherjee2022Zenodo}. We reduced several gridpoints of the post-processed cloudy \texttt{Virga} grid. In the cloudy grid we use here, we included only one heat redistribution factor (0.5), only one intrinsic temperature (100 K), only $f_{sed}$ values $\le$ 3, and only $\log_{10}K_\textrm{zz} > 5$, as this low of a $\log_{10}K_\textrm{zz}$ is unphysically small at temperatures $>$ 500 K  \citeApp[e.g., Fig. 2;][]{moses2022}, as in the atmosphere of WASP-39b. The original grids in ref.\ \citeApp{Mukherjee2022Zenodo} were only computed for wavelengths from 0.3 to 6 $\mu$m; here we extend the simulated transmission spectra of the grid out to wavelengths of 15 $\mu$m.

To assess the presence of SO$_2$ in the MIRI LRS data, we first inject a constant abundance of SO$_2$ into each model at gridpoints of 3 ppm, 5 ppm, 7.5 ppm, 10 ppm, 20 ppm, and 100 ppm, and we then recompute the model spectra. These values of SO$_2$ are therefore not chemically consistent with the rest of the atmosphere. As in the IDIC grid, we fit each transmission spectrum reduction by binning the model spectra (resampled to opacities at R=20,000 \citeApp{batalha2020}) to the resolution of the observations, allow for a vertical offset, and calculate $\chi^2$ for each model spectrum. We take the top 20 best-fitting models to account for scatter in the preferred grid values and discard clear outliers.

Without SO$_2$, although we find comparable overall fits ($\chi^2$ $\le$ 2.6) to the data for the \texttt{Eureka!} reduction, none of the SO$_2$-free RCTE models capture the rise around 7.7 $\mu$m or 8.5 $\mu$m. Once SO$_2$ is added, we find that the overall model fit to the \texttt{Eureka!} reduction is slightly worse ($\chi^2$ $\le$ 2.7), but the shape of the spectrum better matches at 7.7 $\mu$m and 8.5 $\mu$m. This slightly worse fit is driven by the slightly higher transit depths from 5 -- 6 $\mu$m in the \texttt{Eureka!} reduction, which results in a higher baseline ``continuum'' when SO$_2$ is not included. For both the SPARTA and \texttt{Tiberius} reductions, the grid model fits improve with added SO$_2$. Most crucially, in the absence of SO$_2$, the best-fitting clear \texttt{PICASO 3.0} and cloudy \texttt{PICASO 3.0} + \texttt{Virga} grid models across all reductions are dominated by H$_2$O absorption, as well as prominent contributions from CH$_4$ for the \texttt{Tiberius} and \texttt{Eureka!} data, as shown in Extended Data Figure 3. For the \texttt{Tiberius} and \texttt{Eureka!} reductions, cloudy cases without SO$_2$ result in high inferred amounts of CH$_4$ (VMR $\sim$ 1--50\,ppm) at 10 mbar---where the MIRI/LRS observations probe. These CH$_4$ mixing ratios are in disagreement with the lack of CH$_4$ in WASP-39b's atmosphere observed at shorter wavelengths with NIRISS, NIRSpec, and NIRCam (with best-fit models having CH$_4$ VMRs of $\sim$3 ppb, $\sim$0.1 ppm, and $\sim$50 ppb, respectively) \citeApp{feinstein2023,Alderson2023,ahrer2023,rustamkulov2023}. With the \texttt{SPARTA} reduction, rather than compensating for the lack of SO$_2$ opacity with elevated CH$_4$ abundances, the \texttt{PICASO} grid best-fits invoke opacity from a high altitude, optically thick silicate cloud.

Models with SO$_2$ injected produce better overall fits to each MIRI reduction, with mixing ratios of carbon, oxygen, and sulfur-bearing species in agreement with those inferred from shorter wavelength data from NIRISS, NIRSpec, and NIRCam. Therefore, our results suggest MIRI data alone can independently constrain relevant atmospheric gaseous species. With these MIRI data in addition to the previous JWST observations, we demonstrate that SO$_2$ in WASP-39b's atmosphere is required to self-consistently interpret the data from JWST over a wide wavelength range.

When SO$_2$ is included in the RCTE \texttt{PICASO 3.0} models, we find that all three reductions prefer C/O ratios of $\le$ Solar values. These low C/O ratios result from the lack of methane needed to fit the data. Metallicity values range from $\sim$10$\times$ Solar for the \texttt{Eureka!} and \texttt{Tiberius} reductions to $\sim$10-30$\times$ Solar for the SPARTA reduction.  Best-fits are comparable between clear and cloudy cases, with high best-fitting values of $f_{sed}$ resulting in cloud decks below the atmospheric regions probed by MIRI/LRS. The best-fitting models using MIRI therefore result in very different cloud parameters than models fit to shorter wavelengths \citeApp{feinstein2023,Alderson2023,ahrer2023,rustamkulov2023}. These cloud parameter discrepancies highlight that constraining cloud conditions requires wide wavelength coverage and may result from cloud formation localized to different atmospheric layers \citep{miles2023}.

Finally, within the framework of injected uniform \ce{SO2} abundances that do not vary with altitude, we find that all of our \ce{SO2} abundance grid points result in comparable model fits, preventing a strong \ce{SO2} abundance constraint from the \texttt{PICASO 3.0} grid.  

\subsection*{Retrieval Modelling}
In addition to forward modelling, we further investigated the atmosphere of WASP-39b as seen by MIRI/LRS using six different free-retrieval frameworks (see descriptions below). Free retrievals use parameterized atmospheric models to directly extract constraints on atmospheric properties from the data. Each chemical species in the model is treated as an independent free parameter, rather than abundances being calculated under assumptions such as chemical equilibrium or photochemistry. The retrievals presented in this paper all assume that the atmosphere is well-mixed, so chemical abundances are held constant throughout the atmosphere. All retrievals also assume an isothermal temperature profile, since the MIRI-LRS spectrum probes a relatively small range of atmospheric pressures and therefore is relatively insensitive to the temperature structure. All retrievals contain some prescription for aerosols, but the details vary across the six frameworks and are described in more detail below. This variation in aerosol treatment is intentional, and by this approach we hope to capture the impact of different retrieval choices on molecular detection and abundance measurements for MIRI. All frameworks also retrieve either a reference pressure or reference radius, to account for the so-called `normalization degeneracy' (see \citeApp{hengkitzmann17}). Helios-r2 also includes the stellar radius and log($g$), where $g$ is gravitational acceleration, as free parameters. For all frameworks, we ran the preferred model set up, and those removing \ce{H2O} or \ce{SO2}, allowing us to calculate their Bayesian evidence following \citeApp{08Trotta} (Extended Data Table 4).

Atmospheric models do not provide as good a match to the data at $\gtrsim 10 \mu$m, with worse fits by $\chi^2$ and p-value metrics than when only considering data bluewards of 10$\mu$m. Therefore, we considered the possibility of retrieving only on the short wavelengths. While we find that the retrieved abundances are highly sensitive to the wavelengths considered, there is no evident, data-driven argument to disregard data at longer wavelengths, and the fits are acceptable. Therefore, the atmospheric inferences presented below consider the entire MIRI-LRS spectrum from 5 to 12\,$\mu$m. Further investigation into the apparent decrease in transit depth at 10 $\mu$m is warranted in future work.

\subsubsection*{ARCiS}

ARCiS (ARtful modelling Code for exoplanet Science) is an atmospheric modelling and Bayesian retrieval package~\citeApp{18OrMi,20MiOrCh}, which ultilises the Multinest~\citeApp{feroz2009} Monte Carlo nested sampling algorithm to sample a parameter space for the region of maximum likelihood. 
ARCiS is capable of both free molecular and constrained chemistry (i.e. assuming thermochemical equilibrium) retrievals, with the latter using GGchem~\citeApp{Woitke2018} for the chemistry. For this work we use a free molecular retrieval with a simple grey, patchy cloud model. This simple model parameterises cloud-top pressure and the degree of cloud coverage (from 0 for completely clear to 1 for completely covered). We explored the use of a variety of molecular species in our retrievals, with the majority of their abundances being unconstrained by the retrieval of this dataset. In particular, we searched for additional photochemical products including SO and \ce{SO3}. The photochemical model of ref.\ \citeApp{Tsai2023} predicts observable amounts of SO but very little \ce{SO3}. We find some weak-to-moderate (2.5~$\sigma$) evidence for SO~\citeApp{SO_ExoMol_theory} and no evidence of \ce{SO3} \citeApp{16UnTeYu}, qualitatively matching the photochemical model predictions. In addition, we find $\sim$3.3$\sigma$ evidence for the presence of a molecule such as SiH~\citeApp{18YuSiLo}, BeH~\citeApp{18DaTeLa}, or NO~\citeApp{HITEMP_NO}. The broad opacity features from these species, however, are indistinguishable from a continuum effect such as haze.

In the absence of other spectral features from these molecules, and because we do not expect SiH, BeH, or NO to be abundant enough($\sim$1000 ppm are required, compared to a maximum of $\sim$10 ppm for \ce{SiH} and fractions of a ppm for \ce{BeH} under the assumption of solar-abundance thermochemical equilibrium~\citeApp{Woitke2018,Lodders2020}), we exclude them in our models. We therefore present a simplified set of molecules, with only H$_2$O~\citeApp{Polyansky2018} and SO$_2$~\citeApp{Underwood2016} included, along with the parameters for the clouds. Combined with isothermal temperature and planetary radius, this totals six free parameters. The reference pressure for the radius is 10~bar. The opacities are k-tables from the ExoMolOP database~\citeApp{2021Chubb}, with the linelists from the ExoMol~\citeApp{exomol2020} or HITEMP~\citeApp{Rothman2010} database as specified. Collision-induced absorption for H$_2$ and He are taken from refs.\ \citeApp{Borysow2001} and \citeApp{Borysow2002}. We use 1000 live points and a sampling efficiency of 0.3 in Multinest. We used a value of 0.281 $M_{J}$ for the planetary mass, and 0.9324 $R_{\odot}$ for the stellar radius.

\subsubsection*{Aurora}

Aurora is an atmospheric inference framework with applications to transmission spectroscopy of transiting exoplanets \citeApp[e.g.,][]{Welbanks2022,Mikal2023}. The comprehensive description of the framework and modelling paradigm are explained in ref.\ \citeApp{Welbanks2021}. For this dataset we considered a series of atmospheric models ranging from simple cloud-free isothermal models, to those with multiple chemical species, inhomogeneous cloud and hazes, and non-isothermal pressure-temperature (PT) profiles. The parameter estimation was performed using the nested sampling algorithm \citeApp{skilling04} through MultiNest \citeApp{feroz2009} using the PyMultinest implementation \citeApp{buchner2014}. 

We find that the retrieved abundances of H$_2$O and SO$_2$ vary by several orders of magnitude depending on the data reduction considered, the wavelength range included (e.g., above or below 10\,$\mu$m), and assumptions about the atmospheric model used \citeApp[e.g, cloud-free vs. cloudy, fully cloudy vs. inhomogeneous clouds, multiple absorbers vs. limited absorbers; see, e.g.,][]{Welbanks2019}. 

Our initial exploration of atmospheric models finds that when considering multiple species (e.g., Na, K, CH$_4$, NH$_3$, HCN, CO, CO$_2$, C$_2$H$_2$), their abundances are largely unconstrained despite affecting the retrieved SO$_2$ abundances by at least an order of magnitude, generally skewing them towards lower values (e.g., $\log_{10}(\mathrm{SO}_2)\lesssim -6$). The use of parametric PT profiles \citeApp[e.g.,][]{MadhusudhanSeager2009apjRetrieval} do not result in significant changes to the retrieved abundances and the resulting temperature profiles are largely consistent with isothermal atmospheres. Finally, we find that assuming cloud-free or homogeneous cloud cover can result in artificially tight constraints on the H$_2$O abundances as expected \citeApp[e.g.,][]{Welbanks2019, Welbanks2021, Barstow2020a}, motivating our choice to consider the presence of inhomogeneous clouds/hazes. 

Given the above considerations, we settled on a simplified fiducial model to calculate the model preference \citeApp[i.e., `detection'; see, e.g.,][]{Benneke2013, Welbanks2021} for H$_2$O and SO$_2$ with the caveat that the retrieved abundances are highly dependent on the model/data assumptions. This simplified model only considers absorption due to H$_2$O and SO$_2$ using line lists from \citeApp{Rothman2010} and \citeApp{Underwood2016} respectively, H$_2$--H$_2$ and H$_2$--He collision-induced absorption with line lists from \citeApp{Richard2012}, the presence of inhomogeneous clouds and hazes following the single sector model in ref.\ \citeApp{Welbanks2021} \citeApp[see also][]{macdonald2017, Barstow2020a}, and an isothermal pressure temperature profile. In total, our atmospheric model has eight free parameters: two for the constant-with-height volume mixing ratios of the chemical species considered, one for the isothermal temperature of the atmosphere, four for the inhomogeneous clouds and hazes, and one for the reference pressure for the assumed planet radius ($R_{\mathrm{p}}=1.279~\mathrm{R}_{\mathrm{J}}$, $\log_{10}(g)=2.63$~cgs, $R_{\mathrm{star}}=0.932R_{\odot}$). The forward models for the parameter estimation were calculated at a constant resolution R$=10,000$ using 1000 live points for MultiNest. 

\subsubsection*{CHIMERA}
CHIMERA \citeApp{Line2013} is an open-source radiative transfer and retrieval framework which has been extensively used to study the atmospheres of planetary mass objects, ranging from brown dwarfs \citeApp{Line2017} to terrestrial planets \citeApp{May2021}. The forward model is coupled to a nested sampler, namely MultiNest \citeApp{feroz2009} using the PyMultiNest \citeApp{buchner2014} wrapper. CHIMERA takes advantage of the correlated-k approximation \citeApp{lacisoinas,Molliere2015} in order to rapidly compute the  transmission through the atmosphere. Given the flexible nature of the code, it is capable of modelling a range of different aerosol and cloud scenarios \citeApp{Mai2019}, as well as a range of different thermal structures \citeApp{MadhusudhanSeager2009apjRetrieval,ParmentierGuillot2014aapTmodel}.

For this work we are limited to the spectral bands we have access to, thus, we only model H$_2$O and SO$_2$ using line data from refs.\ \citeApp{Polyansky2018} and \citeApp{Underwood2016} respectively. We assume the atmosphere is dominated by H$_2$, with a He/H$_2$ ratio of 0.1764; therefore, we also model the H$_2$--H$_2$ and H$_2$--He collision-induced absorption \citeApp[][]{Richard2012}. We model hazes following the prescription of \citeApp{Lecavelier2008}, which treats hazes as enhanced H$_2$ Rayleigh scattering with a free power-law slope. Alongside the haze calculation, we fit for a constant-in-wavelength grey cloud with opacity $\kappa_\text{cloud}$. We also assess the patchiness of the cloud by linearly combining a cloud-free model with the cloudy model \citeApp{Line2016}. We find that the inclusion of hazes does not improve any of our inferences, thus our final model presented is from using the grey cloud alone. We used a value of 0.281 $M_{J}$ for the planetary mass, and 0.932 $R_{\odot}$ for the stellar radius.

\subsubsection*{Helios-r2}

\texttt{Helios-r2} (The open-source \texttt{Helios-r2} code can be found here: \url{https://github.com/exoclime/Helios-r2}) \citeApp{Kitzmann2020ApJ...890..174K} is an open-source, GPU-accelerated retrieval code for atmospheres of exoplanets and brown dwarfs and can be used for transmission, emission, and secondary-eclipse observations (see, e.g., \citeApp{Bourrier2020A&A...637A..36B}, \citeApp{Mesa2020MNRAS.495.4279M}, or \citeApp{Lueber2022ApJ...930..136L}). It uses a Bayesian nested sampling approach to compute the posterior distributions and Bayesian evidences, based on the \texttt{MultiNest} library \citeApp{feroz2009}.

In \texttt{Helios-r2} the chemical composition can be constrained assuming chemical equilibrium using the \texttt{FastChem} (The open-source \texttt{FastChem} code can be found here: \url{https://github.com/exoclime/FastChem}) chemistry code \citeApp{Stock2018MNRAS.479..865S, Stock2022MNRAS.517.4070S} or by performing a 
free abundance retrieval with either isoprofiles or vertically varying abundances. The temperature profile can also be either described by an isoprofile or allowed to vary with height by using a flexible description based on piece-wise polynomials or a cubic spline approach. Given the limited number of available observational data points in this study, we chose to describe the temperature and the chemical abundances with isoprofiles.

In our final retrieval calculations only two gas-phase species are directly retrieved (\ce{H2O} and \ce{SO2}), while \ce{H2} and He are assumed to form the background atmosphere based on their solar H/He ratio. Additional chemical species, such as HCN, CO, \ce{CO2}, or \ce{CH4} for example, were tested but resulted in unconstrained posteriors. 

We used the Exomol POKAZATEL line list for H$_2$O \citeApp{Polyansky2018} and the ExoAmes SO$_2$ \citeApp{Underwood2016} line list in our retrievals. Line list data for HCN, CO, and \ce{CH4} were taken from \citeApp{HarrisEtal2006}, \citeApp{Li2015}, and \citeApp{Yurchenko2017} respectively.
The opacities were calculated with the open-source opacity calculator \texttt{HELIOS-K} (The open-source \texttt{HELIOS-K} code can be found here: \url{https://github.com/exoclime/HELIOS-K}) \citeApp{Grimm2015ApJ...808..182G, Grimm2021ApJS..253...30G} and are available on the DACE platform (\url{https://dace.unige.ch}). The collision-induced absorption of \ce{H2}--\ce{H2} and \ce{H2}--He pairs was taken from \citeApp{Abel2011}, \citeApp{Abel2012}, and \citeApp{Fletcher2018}.

In the retrieval calculations, we added a grey cloud layer with the cloud's top pressure as a free parameter. Additionally, we used the surface gravity and the stellar radius as free parameters with Gaussian priors based on their measured values to incorporate their uncertainties in the retrieval results.

For the retrieval calculations in this study, 2000 live points and a sampling efficiency of 0.3 for an accurate determination of the Bayesian evidence were used.

\subsubsection*{NEMESIS}
NEMESIS \citeApp{Irwin2008} is an open-source retrieval algorithm that allows simulation of a range of planetary and substellar bodies, using either nested sampling \citeApp{kt2018,skilling04} or optimal estimation \citeApp{rodgers2000} to iterate towards a solution. It has been used extensively to model the atmospheres of transiting exoplanets (e.g., \citeApp{Barstow2020a}). NEMESIS uses the correlated-k approximation \citeApp{lacisoinas} to allow rapid calculation of the forward model. It allows flexible parameterization of aerosols and gas abundance profiles, and can also be used to simultaneously and consistently model multiple planetary phases (e.g., \citeApp{Irwin2020}). 

In this work, we use the nested sampling algorithm PyMultiNest \citeApp{buchner2014,feroz2009}, with 2000 live points. We include H$_2$O line data from the POKAZATEL linelist \citeApp{Polyansky2018} and SO$_2$ line data from the ExoAmes linelist \citeApp{Underwood2016}, using k-tables calculated as in \citeApp{2021Chubb}. Collision-induced absorption information for H$_2$ and He is taken from \citeApp{Borysow2001} and \citeApp{Borysow2002}. Aerosol is modelled as an opaque grey cloud deck, with a variable top pressure. We also retrieve a fractional cloud coverage parameter, simulating the total terminator spectrum as a linear combination of a cloudy spectrum and an otherwise identical clear spectrum. We also tested the inclusion of a simple haze model with a tunable scattering index parameter, after refs.\ \citeApp{macdonald2017} and \citeApp{Barstow2020a}, but found that the retrieved scattering index gave an unrealistically steep spectral slope. We therefore present the models including only a grey cloud deck. We used a value of 0.281 $M_{J}$ for the planetary mass, and 0.9324 $R_{\odot}$ for the stellar radius.

\subsubsection*{PyratBay}

\textsc{PyratBay}\citep{PyratBay-CubillosBlecic2021-docs}, PYthon RAdiative-Transfer in a BAYesian framework, is an open-source software that enables atmospheric forward and retrieval modelling of exoplanetary spectra \citeApp{CubillosBlecic2021-PyratBay}. This software utilizes parametric temperature, composition, and altitude profiles as a function of pressure to generate emission and transmission spectra. The radiative transfer module considers various sources of opacity, including alkali lines \citeApp{BurrowsEtal2000apjBDspectra}, Rayleigh scattering \citeApp{Kurucz1970saorsAtlas, LecavelierEtal2008aaRayleighHD189733b}, Exomol and HITEMP molecular line lists \citeApp{TennysonEtal2016jmsExomol, Rothman2010}, collision-induced absorption \citeApp{Borysow2001, Borysow2002}, and cloud opacities. To optimize retrieval, \textsc{PyratBay} compresses these large databases while retaining essential information from dominant line transitions, using the method described in ref.\ \citeApp{Cubillos2017apjCompress}. The software offers various cloud condensate prescriptions, including the classic ``power law+gray'' model, a ``single-particle-size'' haze profile, a ``patchy clouds'' model with partial coverage factor \citeApp{LineParmentier2016-patchy}, and a complex parameterized Mie-scattering thermal stability model \citeApp{BlecicEtal2023-TSC, KilpatrickEtal2018apjWASP63bWFC3, VenotEtal2020-JWST-WASP-43b}. Furthermore, \textsc{PyratBay} allows users to adjust the complexity of the compositional model, ranging from a ``free retrieval'' approach where molecular abundances are freely parameterized to a ``chemically consistent" retrieval that assumes chemical equilibrium. For the chemically consistent retrieval, users can choose between the numerical TEA code \citeApp{BlecicEtal2016apsjTEA, TEA-docs-Blecic2017} and the analytical RATE code \citeApp{CubillosEtal2019apjRate}, both of which can rapidly calculate volume mixing ratios of desired elemental and molecular abundances across a wide range of chemical species. The software also provides a variety of temperature models, including isothermal profiles and physically motivated parameterized models \citeApp[e.g.,][]{ParmentierGuillot2014aapTmodel, MadhusudhanSeager2009apjRetrieval}. To sample the parameter space and perform Bayesian inference, \textsc{PyratBay} is equipped with two Bayesian samplers: the differential-evolution Markov Chain Monte Carlo (MCMC) algorithm \citeApp{terBraak2008SnookerDEMC}, implemented via ref.\ \citeApp{CubillosEtal2017apjRednoise}, and the nested-sampling algorithm, implemented via PyMultiNest \citeApp{feroz2009, buchner2014}. These algorithms utilize millions of models and thousands of live points to explore the parameter space effectively.

For this analysis, we conducted a free retrieval and tested various model assumptions. These involved testing all temperature parametrizations implemented in our modelling framework, a wide range of chemical species opacities expected to exhibit observable spectral features in the MIRI wavelength region, H$_2$O \citeApp{Polyansky2018}, CH$_4$ \citeApp{HargreavesEtal2020}, NH$_3$ \citeApp{YurchenkoEtal2011, YurchenkoEtal2015}, HCN \citeApp{HarrisEtal2006, HarrisEtal2008}, CO \citeApp{Li2015}, CO$_2$ \citeApp{Rothman2010}, C$_2$H$_2$ \citeApp{WILZEWSKI2016193}, SO$_2$ \citeApp{Underwood2016}, H$_2$S \citeApp{AzzamEtal2016-H2S}, and different cloud prescriptions. Our transmission spectrum was generated at a resolution of R$\sim$15000 and then convolved to match the MIRI resolution of 100. We assumed a hydrogen-dominated atmosphere with a He/H${_2}$ ratio of 0.1764 and accounted for H${_2}$--H${_2}$ \citeApp{Borysow2001} and H${_2}$--He \citeApp{BorysowEtal1989apjH2HeRVRT} collision-induced absorptions. We used the same values of the stellar radius and planetary mass as the NEMESIS pipeline. To evaluate the likelihood of our models, we utilized the PyMultiNest algorithm with 2000 live points. Similar to the findings of other retrieval frameworks, the majority of the considered species were largely unconstrained. The Mie-scattering cloud models did not detect spectral signatures of any condensates in the data, and the more complex temperature models yielded temperature profiles that were largely consistent with an isothermal atmosphere. Only H$_{2}$O and SO$_{2}$ exhibited detectable spectral features in the data, and the assumption of a patchy gray cloud was the most suitable for the quality of the observations. Our final atmospheric model, applied to each team's reduction data, consisted of six free parameters: two for the constant-with-height volume mixing ratios of the chemical species, one for the isothermal temperature of the atmosphere, one for the planetary radius, and two for the patchy opaque cloud deck.

\subsubsection*{TauREx}

\texttt{TauREx}, Tau Retrieval for Exoplanets, is an open-source fully Bayesian inverse atmospheric retrieval framework \citeApp{Waldmann2015a,Waldmann2015b}. We adopted the latest version (3.1) of the \texttt{TauREx} software \citeApp{Al-Refaie2021,Al-Refaie2022}. This version makes exclusive use of absorption cross sections, as the correlated-$k$ tables are no longer computationally advantageous \citeApp{Al-Refaie2021}. We selected the PyMultinest algorithm to sample the parameter space \citeApp{feroz2009,buchner2014}. The atmosphere was modeled with 200 equally spaced layers in log-pressure between 10$^6$ and 10$^{-4}$\,Pa. In all our tests, we assumed an isothermal profile and constant mixing ratios with altitude. The radiative transfer model accounts for absorption from chemical species, collision-induced absorption by H$_2$--H$_2$ and H$_2$--He \citeApp{Abel2011,Abel2012,Fletcher2018}, and clouds. We performed initial retrieval tests including a long list of molecular species, H$_2$O \citeApp{Polyansky2018}, SO$_2$ \citeApp{Underwood2016}, CO \citeApp{Li2015}, CO$_2$ \citeApp{Rothman2010}, CH$_4$ \citeApp{Yurchenko2017}, HCN \citeApp{Barber2014}, NH$_3$ \citeApp{Coles2019}, FeH \citeApp{Wende2010} and H$_2$S \citeApp{AzzamEtal2016-H2S}, but found that only H$_2$O and SO$_2$ may have detectable features in the observed MIRI spectra. We validated statistically the detection of both H$_2$O and SO$_2$ by comparing the Bayesian evidence of best-fit retrievals with both species versus those obtained by removing either molecule. We considered the following scenarios: (1) a clear atmosphere, (2) an atmosphere with an optically-thick cloud deck, for which we fitted the top-layer pressure, and (3) an atmosphere with haze, using the formalism of ref.\ \citeApp{Lee2013} for modelling the Mie scattering. We finally selected the retrievals with a thick cloud deck, which provide the most consistent scenarios across data reductions, and with slightly more conservative error bars. Only for the \texttt{Eureka!} reduction, the haze model was slightly favored (2.4$\sigma$), but the corresponding molecular abundances are affected by strong degeneracy between water and haze. For other reductions, the inferred molecular abundances are essentially independent of the retrieval scenario.
We used a value of 0.281 $M_{J}$ for the planetary mass, and 0.939 $R_{\odot}$ for the stellar radius.

\subsubsection*{Free retrieval results}

The results from all retrieval frameworks, across all three reductions, are presented in Extended Data Table 4 and shown in Extended Data Figure 4. These serve to illustrate the general consistency of the results for \ce{SO2} and \ce{H2O}, whilst also highlighting the differences in retrieved abundance for some cases. We reiterate that the different retrieval teams made a variety of choices in the setup of their retrievals, which are described in more detail above. The overall good agreement is testament to the robustness of our detection of \ce{SO2} in the MIRI dataset. 

We recover a range of median abundances for log(\ce{SO2}) between $-5.9$ and $-5.0$ across all reductions and retrieval frameworks.
The overall spread of log(\ce{SO2}) across all retrievals and reductions, from the lowest $-1\sigma$ bound to the highest $+1\sigma$ bound, is $-6.4$---$-4.6$ (the range reported in the main text refers only to the retrievals on the \texttt{Eureka!} reduction), corresponding to volume mixing ratios of 0.4---25 ppm (0.5---25 ppm if only retrievals on the \texttt{Eureka!} reduction are considered). Note that this range could potentially be wider if a more extensive exploration of possible cloud and haze configurations were conducted, which we leave to future work. 

\ce{SO2} is detected at more than 3$\sigma$ significance in all cases except the Helios-r2 retrievals for \texttt{Eureka!} and SPARTA (2.54$\sigma$ and 2.99$\sigma$ respectively), and the Aurora retrieval for SPARTA (2.95$\sigma$). The Helios-r2 model has the simplest representation of clouds, but also allows the stellar radius and planetary log($g$) to vary, so it is likely that the precise combinations of the \texttt{Eureka!} and SPARTA spectra and the chosen variables result in weaker detections for \ce{SO2}, because other parameters have more freedom to compensate for a lack of \ce{SO2} in this framework. Similarly, the Aurora framework has a unique representation of aerosol, including both cloud and haze, with the cloud top pressure as a free parameter. This also increases the flexibility of the model to compensate for changes in the SO$_2$ abundance.  
In summary, free retrievals provide a broadly consistent picture, which is also consistent with the \ce{SO2} volume mixing ratios from the best-fitting photochemical models (see e.g. Figure 4). 

Test runs with the ARCiS retrieval also included SO opacity, which was not included in the other retrieval schemes. The existence of SO is not ruled out by these retrievals, with weak-to-moderate (2.5$\sigma$) evidence for it being present in the atmosphere. If present, it contributes to the spectrum at around 9 $\mu$m and is an additional source of opacity overlapping with the longer wavelength end of the broad \ce{SO2} feature. The presence of SO is consistent with photochemical predictions, and should be an avenue for future exploration. 

We also retrieve log(\ce{H2O}) abundances in all cases. For the most part, the median values for nearly all retrievals and reductions range from log(\ce{H2O}) of -2.3 to -1.1, with an anomalously low value for the \texttt{Eureka!} reduction and the Aurora (-3.9) retrieval. This retrieval framework includes haze, so we postulate that in this case the haze slope is compensating for the shape of the \ce{H2O} feature. Whilst the CHIMERA retrieval also includes haze and cloud, the cloud is uniformly distributed and the opacity is scaled, whereas Aurora has the cloud top pressure as a free parameter. This likely accounts for the different solutions between these two codes.  The \texttt{Eureka!} reduction also results in a spectrum with a slightly smoother downward slope between 5.2 and 6.5 $\mu$m than the other two reductions, which contributes to the preference for haze over H$_2$O absorption in the Aurora retrieval.

The main \ce{H2O} absorption feature in the MIRI-LRS range is a broad feature centered around 6\,$\rm{\mu}$m, but extending beyond the short wavelength cut off and also into the region affected by \ce{SO2}. Slight differences in the shape of the spectrum between the three reductions at the shortest wavelengths, which is the region most sensitive to \ce{H2O}, drive the subtle differences in the retrieved \ce{H2O} abundances between those reductions.  \texttt{Eureka!} and SPARTA have very similar transit depths and yield slightly larger \ce{H2O} abundances (range excepting outliers: -1.9 to -1.1) than the Tiberius reduction (range: -2.3 to -1.5).

Whilst all retrievals include some prescription for cloud and/or haze, the parameters are generally poorly constrained. For ARCiS, CHIMERA and \texttt{PyratBay}, no meaningful constraints on any cloud properties were obtained for any reductions. For \texttt{Helios-r2}, 1$\sigma$ lower limits on log(cloud top pressure) in bar of -1.85, -1.62, and -1.78 are found for the \texttt{Eureka!}, \texttt{Tiberius}, and SPARTA reductions respectively. Similarly, \texttt{TauREx} provides 1$\sigma$ lower limits on log(cloud top pressure) of -1.60, -1.97 and -2.03 for \texttt{Eureka!}, \texttt{Tiberius} and SPARTA. For NEMESIS, we find that the cloud top pressure and cloud fraction are degenerate, but high cloud fractions with low cloud top pressures are not permitted, so we can rule out high, opaque cloud covering a large percentage of the terminator. 
For Aurora/\texttt{Eureka!}, the haze scattering slope is constrained to $\gamma$ = $-4.6 ^{+ 1.0 }_{- 1.8 }$, consistent with a Rayleigh-scattering slope ($\gamma$ = -4) within 1-$\sigma$. 
In summary, we can rule out a grey cloud extending to low pressures with broad terminator coverage, but otherwise with such varied results across reductions and retrievals we cannot place any constraints on cloud or haze properties.

\bmhead{Data Availability}
The data used in this paper are associated with JWST program DD-2783 and are available from the Mikulski Archive for Space Telescopes (\url{https://mast.stsci.edu}). The data products required to generate Figs. 1-4 and Extended Data Figs. 1-5 are available here: https://doi.org/10.5281/zenodo.10055845. All additional data are available upon request.

\bmhead{Code Availability} 
\begin{flushleft}
The codes VULCAN and gCMCRT used in this work to simulate composition and produce synthetic spectra are publicly available:
VULCAN\textsuperscript{\citeApp{tsai17,Tsai2021}} (\url{https://github.com/exoclime/VULCAN})\\
gCMCRT\textsuperscript{\citeApp{Lee2022}} (\url{https://github.com/ELeeAstro/gCMCRT})\\
The \texttt{SPARTA} software to reduce JWST MIRI and NIRCam time-series spectra is publicly available: \texttt{SPARTA}\textsuperscript{\citeApp{kempton_2023}}(\url{https://github.com/ideasrule/sparta}).
The \texttt{Tiberius} software to reduce and analyse JWST time-series spectra is publicly available:
\texttt{Tiberius}\textsuperscript{\citeApp{Kirk2017, 2021AJ....162...34K}}(\url{https://github.com/JamesKirk11/Tiberius}).
Six of the free retrieval codes are available at the following locations: 
ARCiS (\url{https://github.com/michielmin/ARCiS}); 
CHIMERA (\url{https://github.com/mrline/CHIMERA}); 
Helios-r2 (\url{https://github.com/exoclime/Helios-r2}); 
NEMESIS (\url{https://github.com/nemesiscode/radtrancode}); 
PyratBay (\url{https://github.com/pcubillos/pyratbay}); 
TauREx (\url{https://github.com/ucl-exoplanets/TauREx3_public}). 

The \texttt{Eureka!} analyses used the following publicly available codes to process, extract, reduce and analyse the data: STScI's JWST Calibration pipeline \citeApp{jwst_v1.8.2}, \texttt{Eureka!} \citeApp{bell2022}, starry \citeApp{starry}, PyMC3 \citeApp{pymc3}, and the standard Python libraries numpy \citeApp{numpy}, astropy \citeApp{astropy2013, astropy2018}, and matplotlib \citeApp{matplotlib}.

\end{flushleft}

\bmhead{Acknowledgments} 
This work is based on observations made with the NASA/ESA/CSA JWST. The data were obtained from the Mikulski Archive for Space Telescopes at the Space Telescope Science Institute, which is operated by the Association of Universities for Research in Astronomy, Inc., under NASA contract no. NAS 5-03127 for JWST. These observations are associated with program no. JWST-DD-2783, support for which was provided by NASA through a grant from the Space Telescope Science Institute. T.B.~acknowledges funding support from the NASA Next Generation Space Telescope Flight Investigations program (now JWST) via WBS 411672.07.05.05.03.02. J.K.B. is supported by a UKRI/STFC Ernest Rutherford Fellowship (grant ST/T004479/1). J.T is supported by the Eric and Wendy Schmidt AI in Science Postdoctoral Fellowship, a Schmidt Futures program. J.B. acknowledges the support received in part from the NYUAD IT High Performance Computing resources, services, and staff expertise. G.M. has received funding from the European Union's Horizon 2020 research and innovation programme under the Marie Sk\l{}odowska-Curie grant agreement No. 895525, and from the Ariel Postdoctoral Fellowship program of the Swedish National Space Agency (SNSA). B.-O. D. acknowledges support from the Swiss State Secretariat for Education, Research and Innovation (SERI) under contract number MB22.00046. E.A.M. acknowledges support from the Centre for Space and Habitability (CSH) and the NCCR PlanetS supported by the Swiss National Science Foundation under grants 51NF40\_182901 and 51NF40\_205606. We thank M. Marley for constructive comments.

\bmhead{Author Contribution}
All authors played a significant role in one or more of the following: development of the original ERS proposal, development of the DDT proposal, preparatory work, management of the project, definition of the observation plan, analysis of the data, theoretical modelling, and preparation of this paper. Some specific contributions are listed as follows: D.P., E.K.H.L., J.L.B., P.G., S.-M.T., V.P., X.Z., J.K.B., J.T., J.K., M.L.-M., and K.B.S. made significant contributions to the design of the program. D.P., A.D.F., and P.G. provided overall program leadership and management. T.B., J.K., and M.Z. reduced the data, modelled the light curves, produced the planetary spectrum, and compared the different data analyses. J.T. and J.K.B. provided free retrieval analyses and also led the free retrieval efforts. S.-M.T. provided a forward model fit to the data and also led the forward modeling efforts. J.B., K.L.C., D.K., G.M., and L.W. provided free retrieval analyses. S.E.M. and I.J.M.C. contributed extensive forward model grids for constraining atmospheric metallicity and elemental ratios. S.J., J.I.M., and J.Y. contributed forward models, which were post-processed into spectra by E.K.H.L.. E.-M.A., A.B.-A., J.B., N.C., B.-O.D., K.D.J., E.A.M., A.D., R.H., P.-O.L., and J.I. contributed additional data reductions that are not shown in this paper, but provided valuable context for the highlighted reductions that was summarized by T.B.. S.L.C., L.F., M.L.-M., A.A.A.P., B.V.R., M.R., and S.R. served on the red team review of the paper, with J.L.B., R.H., and X.Z. offering additional vital comments. A.D.F., J.T., and S.E.M. generated the figures for this paper. D.P., A.D.F., P.G., J.K.B., T.B., J.K., M.Z., S.-M.T., S.E.M., and I.J.M.C. made significant contributions to the writing of this paper. J.T., J.B., K.L.C., S.J., D.K., G.M., J.I.M., L.W., and J.Y. also contributed to the writing of this paper.

\bmhead{Competing interests} The authors declare no conflict of interests.

\bmhead{Corresponding Author} Correspondance to Diana Powell.

\bibliographystyleApp{sn-standardnature} 
\bibliographyApp{master_all_bib_update.bib}

\subsection*{Author Affiliations}
~
\end{document}